# Continuous three-dimensional imaging of nanoscale dynamics by *in situ* electron tomography

Timothy M. Craig[1], Adrien Moncomble[1], Ajinkya A. Kadu[1], Gail A. Vinnacombe-Willson[2,3], Luis M. Liz-Marzán[2,3,4,5], Robin Girod[1,*] and Sara Bals[1,*]

[1] EMAT and NANOlight Center of Excellence, University of Antwerp, Groenenborgerlaan 171, Antwerp B-2020, Belgium
[2] CIC biomaGUNE, Basque Research and Technology Alliance (BRTA), 20014 Donostia-San Sebastián, Spain
[3] Centro de Investigación Biomédica en Red, Bioingeniería, Biomateriales y Nanomedicina (CIBER-BBN), 20014 Donostia-San Sebastián, Spain
[4] Ikerbasque, Basque Foundation for Science, 48009 Bilbao, Spain
[5] CINBIO, University of Vigo, 36310 Vigo, Spain

* Corresponding authors. Email: sara.bals@uantwerpen.be; robin.girod@uantwerpen.be

## Abstract

Visualizing nanoscale transformations in three dimensions (3D) is essential for understanding materials evolution under operating conditions, yet dynamic electron tomography remains limited by slow tilt series acquisition and by reconstruction methods that assume static specimens. These constraints prevent continuous 3D imaging of evolving structures and require electron doses that can alter the specimens and their dynamics. Here, we demonstrate a framework for dynamic electron tomography that combines continuous tilting with a self-supervised deep-learning reconstruction strategy. Our approach incorporates the temporal aspect into the electron tomography reconstruction process to recover 3D volumes at arbitrary time points from a single tilt series. We validate the method using simulations and demonstrate its merit in experimental studies of heat-induced transformations, including morphological evolution of Au nanostars and alloying in Au@Ag nanocubes. Our results establish a practical framework for dynamic, dose-efficient electron tomography, enabling *in situ* 3D investigation of nanomaterial transformations as well as the characterization of beam-sensitive structures.

## Introduction

Characterizing the three-dimensional (3D) morphology and structure of complex nanomaterials is crucial for understanding their properties, optimizing their performance and developing novel applications.[1,2] At the nanoscale, electron tomography (ET) is one of the most widely adopted approaches for 3D imaging.[3,4] In ET, a tilt series of 2D projection images is collected over a tilt range (e.g., ±70°) at fine angular increments (1° – 3°). The series is then computationally reconstructed into a 3D volume through algorithms such as filtered back projection (FBP) or iterative reconstruction techniques like simultaneous iterative reconstruction tomography (SIRT) and expectation maximization (EM).[3,5,6] Conventional ET is nonetheless a relatively time-consuming technique due to the high number of projections required for each reconstruction. More fundamentally, conventional reconstruction algorithms assume that the specimen remains unchanged during acquisition, which has typically limited its application to static objects.

In many cases however, dynamic events occur during acquisition. A common example is that of electron-beam-induced damage, as the repeated imaging of a specific area of interest and the accordingly accumulated dose may induce shrinkage, sputtering or movement.[7,8] Another case of major interest is that of *in situ* TEM, which enables understanding materials evolution under operating conditions.[9] Using specialized holders that apply stimuli (e.g., heat, biasing, stress)[10,11] or introduce an application-relevant environment (e.g., gas, liquid)[12–14] directly in the microscope, dynamic processes can now be triggered and monitored in real-time. However, any specimen transformation, if happening during tilt series acquisition, will be averaged in the final reconstruction and result in so-called motion blur, reducing the spatial resolution.[15]

Recent advances in acquisition schemes have significantly reduced the time needed for each tilt series during an ET experiment, from approximately an hour to just 3-5 minutes using so-called "fast ET".[16–20] In combination with a temporary halt or quench of the stimuli, a tilt series can be acquired at selected time points, leading to *quasi in situ* 3D characterization.[21] This "stop-and-go" ET approach has already enabled monitoring heat-induced structural or compositional changes of plasmonic nanoparticles,[10,21–24] but meaningful temporal sampling requires numerous heat-quench cycles so that the total duration of the experiment easily exceeds 8 hours. This extended period is logistically challenging, and more importantly, introduces concerns regarding the integrity of the samples. For example, it was found that surface ligands surrounding plasmonic nanoparticles transformed into a protective carbon layer upon prolonged electron beam exposure, preventing nanoparticle deformation at temperatures up to 400 °C, even though *ex situ* results showed that these particles change morphology at much lower temperatures.[18,25–27] Furthermore, some transformations may be sensitive to time delays and not all processes can be easily paused for data acquisition, limiting the ability of "stop-and-go" ET to monitor dynamic transformations under *in situ* conditions. Thus, intermediate interruptions render "stop-and-go" ET experiments challenging and potentially unrealistic, highlighting the need for a more time- and dose-efficient, continuous 3D characterization for *in situ* TEM.

Instead of acquiring many sequential tilt series, one may think of recording a continuous tilt series from which reconstructions can be computed within a moving temporal window consisting of a specific number of projections, e.g., a window of 50 projections sequentially using projections 1-50, 2-51, 3-52, etc. This so-called continuous tomography scheme affords an increased number of 3D reconstructions from the same number of 2D images by sharing projections across multiple reconstructions. In practice, projection sharing requires an approximately constant angular range when subdividing the tilt series chronologically, so that the conventional incremental tilt scheme – ranging, for example, from -70° to +70° in 2° increments – is not compatible.[15] To overcome this limitation, we have recently demonstrated that projection sharing can be achieved in ET with a modified golden ratio scanning (GRS) tilt scheme,[15,28,29] which continuously subdivides the tilt series irrationally according to the golden ratio. Still, continuous acquisition during a dynamic process presents a trade-off between spatial and temporal resolution. Using a larger window (more projections per reconstructions) benefits the signal-to-noise ratio and reduces the angular increments but also increases motion blur artifacts.

In parallel, diverse computational methods have been developed for tomographic reconstruction from an increasingly small number of projections. This is typically achieved by including *a priori* knowledge, for instance by assuming the specimen consist of several discrete, homogeneous materials in discrete algebraic reconstruction (DART) or total-variation minimization (TVM).[6,30–32] More recently, deep learning (DL) has found numerous applications in the context of ET including denoising,[33–35] missing wedge inpainting,[36,37] and reconstruction refinement.[38,39] Yet, training of DL models often relies on extensive training datasets of similar specimen or simulated data,[40] oversampling,[36] or subtomogram averaging.[34,35] For the data-scarce regime typical of materials science, self-supervised training schemes, which solely rely on the acquired dataset for training, have therefore been gaining interest. Examples include predicting pixels from their (noisy) spatio-temporal surroundings in denoising tasks,[41,42] or enforcing consistency between the forward projection of a candidate reconstruction and experimental data in ET.[43,44] Implicit neural representations are a particularly suitable DL framework for using such a projection-consistency approach to tomography. They store volumes not as voxelated datasets but directly as a continuous function in a neural network that can be queried at specific coordinates to produce a reconstruction. For data-scarce tomography tasks, neural representations benefit from the implicit bias of neural networks that makes the underlying signal easier to model than noise, and have for instance been shown to yield promising results for denoising and missing wedge compensation in ET.[45] Further, neural representations can also be extended to complete volume-time tomographic series, implicitly regularizing the reconstructions for smoothness in both time and space and making efficient use of all available spatio-temporal information.[46] Yet, despite promising results in magnetic resonance imaging (MRI),[46] X-ray computed tomography (CT),[47–50] and even on simulated electron microscopy data,[51,52] no such neural representation has been effectively applied to experimental dynamic ET data.

Here, we show that dynamic ET experiments can indeed be achieved with self-supervised DL and implicit neural representation when supported by continuous projections acquisition like the GRS tilt scheme. We first present the Deep Image Prior in Space-Time Environment Reconstruction (DIP-STER) network, which is trained solely from a single, continuous tilt series of 2D projection images acquired with the GRS tilt scheme. The method is then evaluated on simulated samples representative of dynamic events including melting and alloying. We finally demonstrate its experimental implementation by continuously tracking the 3D heat-induced evolution of Au nanostars and Au@Ag nanocubes, yielding access to previously elusive nanoscale dynamics.

# Results

## Principles of DIP-STER reconstructions

The general framework to obtain 3D series with high spatio-temporal resolution using DIP-STER is illustrated in **Fig. 1**. We acquire a single tilt series using the GRS tilt scheme that we previously adapted for ET (**Fig. 1a, Supplementary Fig. 1**),[15,29] so that the complete angular range accessible by the specimen holder is continuously sampled, following **Eq. 1**:

$$\theta_i = i(\theta_{max} - \theta_{min})\frac{1+\sqrt{5}}{2}\left(\mathrm{mod}(\theta_{max} - \theta_{min})\right) + \theta_{min} \qquad \textbf{(1)}$$

Where $\theta_i$ is the projection angle at index i, $\theta_{min}$ and $\theta_{max}$ are the lower and upper limit of the tilt range (typically, ± 70°). Thus, each image from the tilt series is taken at a different projection angle and corresponds to a different stage of deformation of the sample (**Fig. 1b**). A DIP-STER neural network is then trained in a self-supervised manner on this tilt series and eventually learns to produce time-dependent 3D reconstructions (**Fig. 1b**).

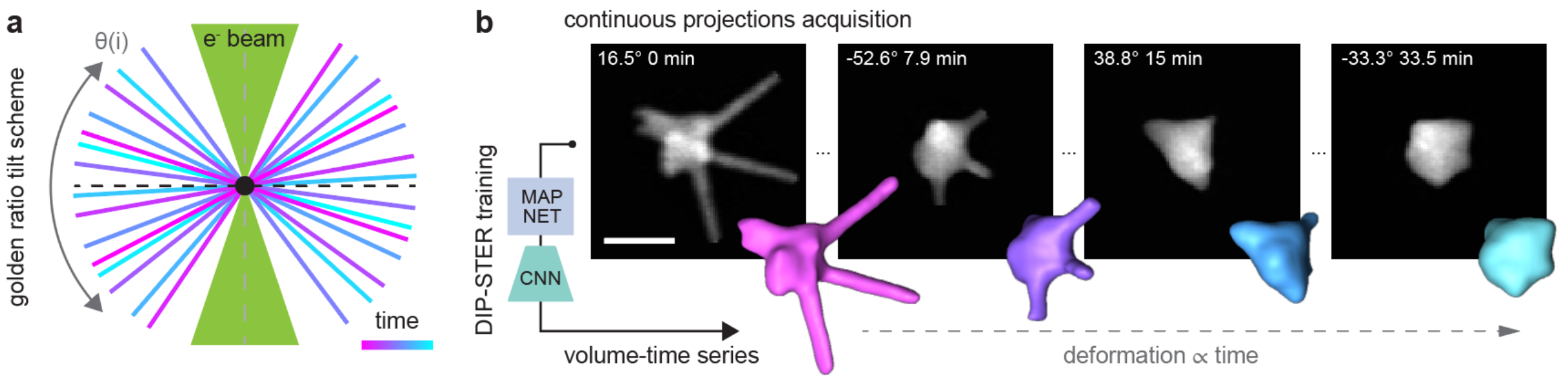

**Fig. 1: DIP-STER reconstruction of Au nanostar heat-induced evolution. a,** Illustration of the GRS tilt scheme which allows continuous tilting and projection acquisition during dynamic experiments such as heating. **b**, DIP-STER is trained in a self-supervised manner on the acquired tilt series and learns to produce time-dependent reconstructions.

3D reconstructions obtained using DIP-STER are fundamentally different from those resulting from conventional ET, as detailed in **Fig. 2**. From a continuous tilt series of an evolving specimen (**Fig. 2a**), a conventional approach would consist in selecting a number of projections to form chronological "sub-tilt series", and using analytical or iterative approaches for tomographic reconstruction (e.g., using 30 projections as input for SIRT, **Fig. 2b**). This amounts however to using a moving-average time window, so that the reconstructions will be blurred and the original state of the specimen remains inaccessible (**Fig. 2b**). In contrast, DIP-STER combines deep image priors (DIP)[43,44,53] and manifold learning[54,55] to recover orthoslices at selected temporal and spatial coordinates (**Fig. 2c**).

A DIP approach solves a minimization problem in which the parameters $\boldsymbol{\varphi}^*$ of a convolutional neural network (CNN) $f_\phi$ are optimized until its output $f_\phi(\mathbf{c})$ agrees with measurements **m**.[46,56,57] The network only takes a latent variable **c** as input, which may be chosen or may be random. During training, the output is transformed back to measurement space via a non-learned forward model *H*, so that $H\left(f_{\boldsymbol{\varphi}}(\mathbf{c})\right)$ is directly comparable to **m** (**Fig. 2c**). The discrepancies between $H\left(f_{\boldsymbol{\varphi}}(\mathbf{c})\right)$ and **m** are evaluated by a loss function $\mathcal{L}$ to serve as the basis for optimization. The method can naturally be expanded to work with a set of N measurements, and the output $f_{\boldsymbol{\varphi}}(\mathbf{c_i})$ is obtained from latent vector $\mathbf{c}_i$ and compared to measurement $\mathbf{m}_i$ in the set. The general optimization function is then:

$$\boldsymbol{\varphi}^* = \operatorname{argmin}_{\boldsymbol{\varphi}} \frac{1}{N} \sum_{i=0}^{N} \mathcal{L}\left(\mathbf{m_i} - \mathrm{H}\left(f_{\boldsymbol{\varphi}}(\mathbf{c_i})\right)\right) \qquad \textbf{(2)}$$

In DIP-STER, and generally for tomography tasks, the set of measurements **m** is the tilt series, and the forward model *H* is the projection operation at angle $\theta_i$. The task is therefore for the network to associate a set of latent variables $\boldsymbol{c}_i$ with reconstructions whose projections agree with the tilt series. It is noteworthy that the operation can be performed in 3D – the output $f_{\boldsymbol{\varphi}}(\mathbf{c_i})$ is a full 3D reconstruction and *H* the 2D projection operator – or in 2D – the output is an orthoslice in the y-direction parallel to the rotation axis whose 1D projection is compared with the corresponding intensities at position y in the experimental tilt series (**Fig. 2c**). Given the computational cost of 3D CNNs,[58] (**Supplementary Note 1**) we use the latter approach and reconstruct 3D volumes layer-by-layer, by stacking reconstructed 2D orthoslices in the y-direction. Measurements $\mathbf{m}_{i,k}$ are therefore indexed by their position i in the tilt series and k along the y-axis.

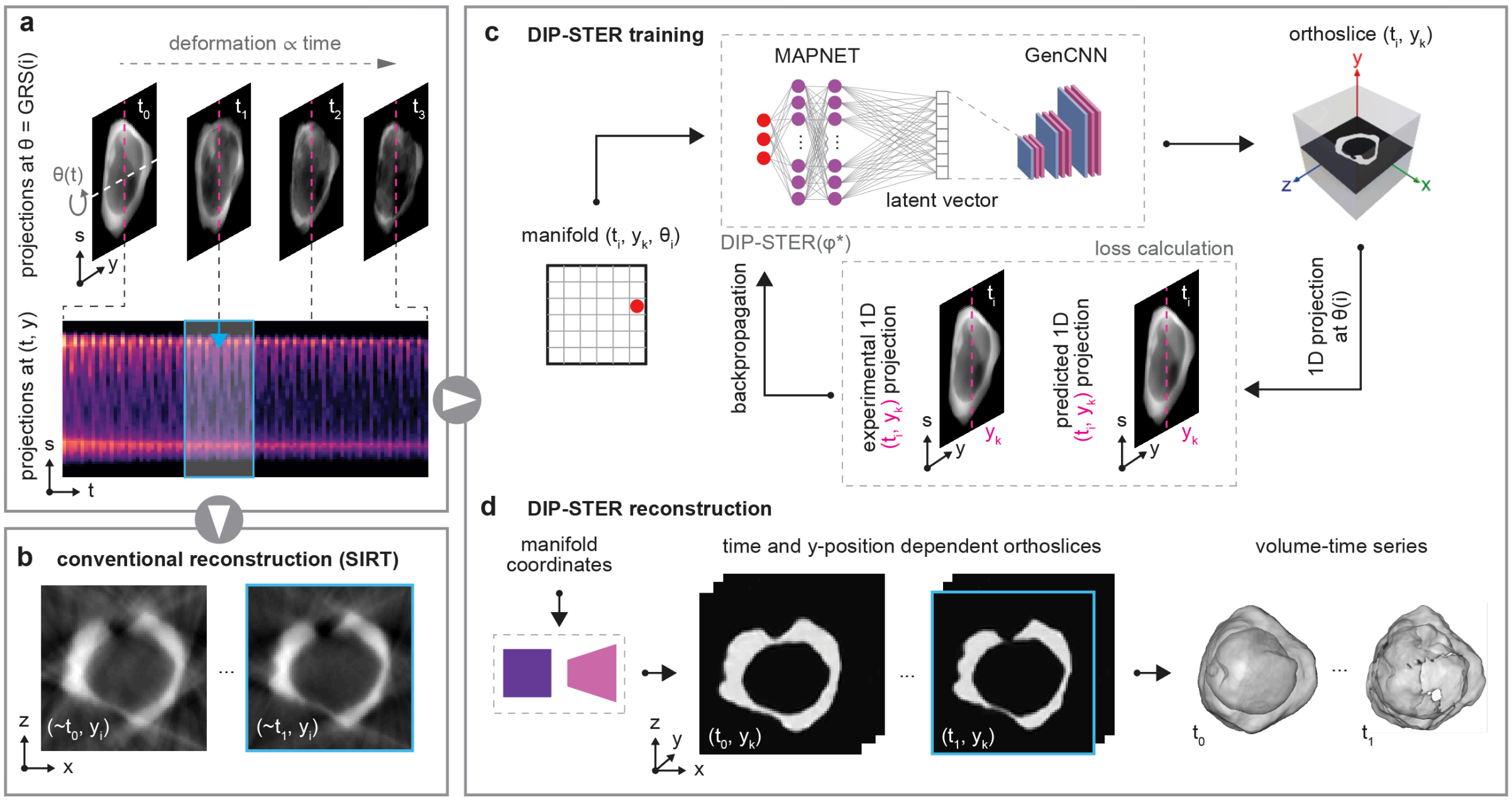

**Fig. 2. Workflow for conventional and DIP-STER dynamic tomographic reconstruction**. **a**, Dynamic tomography starts from a continuous tilt series acquired following the GRS scheme (top). Because the specimen continuously evolves, the 1D projection series can be visualized as a function of time (bottom), where each angle $\theta_i$ corresponds to a time step $t_i$ (blue arrow). The s axis is the position along the detector plane. **b**, Example SIRT reconstructions using 30 projections from the tilt-time series. Each reconstruction necessarily averages across a temporal window (such as the blue rectangle in **a**), resulting in motion blur and large missing wedge artefacts. **c**, DIP-STER training scheme. For each training step, coordinates ($t_i$, $y_k$, $\theta_i$) are randomly selected and passed to an encoder network (MAPNET) to obtain a compressed latent vector. This representation is then passed through a generative CNN to obtain a 2D orthoslice. The loss function is calculated by evaluating the difference between the forward projection of the orthoslice at angle $\theta_i$ and the tilt series slice at ($t_i$, $y_k$) (dashed magenta lines). The error is back-propagated to update the network parameters, and the next step proceeds with different coordinates on the manifold. **d**, After training, the full reconstruction is obtained by passing the coordinates of every 1D tilt series slice. A 3D volume is resolved by stacking 2D orthoslices along the y axis. Temporal dynamics can be reconstructed by repeating this process for every time coordinate.

DIP is also advantageous in that it readily allows learning relationships in time and space by manipulating the latent variable in a manifold learning approach.[46,54,55] In **Eq. 2**, measurements were considered independent, whereas each of them represents a view of the same continuously deforming specimen at time $t_i$ and position $y_k$. Variations in time and y-position in the set of reconstructed orthoslices can therefore be assumed smooth. Such a prior can be injected into the network by sampling inputs **c** not randomly, but on a manifold with axes representing time, y-position and the angle θ of the projections. By setting latent variables as

$$\mathbf{c}_{\mathbf{i,k}} = \left(\frac{y_k}{N^y}, \frac{t_i}{N^t}, \theta_i\right) \qquad \textbf{(3)}$$

and comparing the output with the corresponding projections at $\mathbf{m}_{i,k}$ , the network is encouraged to associate closeness on the manifold with spatial and temporal closeness of the reconstructed orthoslices. This promotes continuity across orthoslices that are close in time and space. To allow for more flexibility to dynamically learn the relationship between projections, we again follow Yoo *et al.* and use a fully-connected mapping network (MAPNET) $g_\psi$ to first convert our 3-coordinates manifold into a more expressive latent vector before going into the CNN.[46] The complete DIP-STER network is detailed in **Supplementary Fig. 2**, and the final optimization function is therefore:

$$\{\boldsymbol{\varphi}^*, \boldsymbol{\psi}^*\} = \operatorname{argmin}_{\boldsymbol{\varphi},\boldsymbol{\psi}} \frac{1}{\mathrm{NK}} \sum_{\mathrm{i}=0}^{\mathrm{N}} \sum_{\mathrm{k}=0}^{\mathrm{K}} \mathcal{L}\left(\mathbf{m}_{\mathbf{i},\mathbf{k}} - \mathrm{H}\left(\mathrm{f}_{\boldsymbol{\varphi}} \circ g_{\boldsymbol{\psi}}(\mathbf{c}_{\mathbf{i},\mathbf{k}})\right)\right) \qquad \textbf{(4)}$$

We use the Geman-McClure loss function due to its resilience to noise and outliers (**Methods**).[59] In summary, the reconstruction of a volume-time series using DIP-STER splits the reconstruction into two phases (**Fig. 2c, d**). During the training phase, random projection coordinates ($t_i$, $y_k$, $\theta_i$) are used as inputs to DIP-STER that reconstructs an orthoslice. The reconstruction is forward projected at angle $\theta_i$ and the resulting projection is compared to the original tilt series slice at the same coordinates to serve as the basis for optimizing the weights of the network (**Fig. 2c**).

After completing the training, a DIP-STER model holds an implicit network representation of an evolving object, with the ability to reconstruct projection-consistent voxelated data for any specified moment in time. This reconstruction is achieved by stacking all the orthoslices, each corresponding to different y-positions and times, and eventually provides volume-time series (**Fig. 2d**). It is worth noting that the continuous nature of the manifold means that *any* coordinates can be queried for reconstruction. Indeed, unless stated otherwise, all results hereafter were produced at coordinates ($t_i$, $y_k$, 0°) during the final inference rather than the ($t_i$, $y_k$, $\theta_i$) training coordinates. We will show later that this approach improves the temporal regularization and guards against outlier projections and residual misalignments in the tilt series.

**DIP-STER evaluation on synthetic data**

To assess the capabilities of DIP-STER for accurately recovering 3D dynamics, we simulated two samples with various deformation types, inspired by real experimental data.[10,22,23] The first case represents an Au nanostar with material redistribution from the branches to the core to simulate morphological evolution under, e.g., heat. To include non-linear and abrupt dynamics, 100 time steps were simulated with only the 33 central ones subject to deformation (**Fig. 3a**, **Methods**). The rate of deformation was set to present a challenging case, with branches mostly disappearing within 15 frames and extensive material movement between frames. The second case simulates internal intensity redistribution with an Au@Ag bimetallic, core-shell nanorod subjected to heating to induce alloying (**Fig. 3f**, **Methods**). **Fig. 3a** and **f** show isosurface renderings and orthoslices at selected time steps from these ground truth datasets. From these dynamic samples, we obtain synthetic tilt-time series by calculating a single projection at an angle set by the GRS formula for each time step (**Supplementary Fig. 3**). Two DIP-STER networks were trained on these tilt-time series, and subsequently used to reconstruct a $128^3$ $vx^3$ volume at each time step. Selected isosurface renderings and orthoslice are displayed in **Fig. 3b** and **g** (additional time steps are in **Supplementary Fig. 4** and **5**). Qualitatively, DIP-STER predictions show high fidelity. Reconstructions of the Au nanostar only evolve within the deformation window, and are stable outside of it (**Fig. 3a** and **b**). For the nanorod with simulated alloying, the boundary between the Au core and Ag shell blurs as the two materials mix, with a sharp core-shell structure at early times.

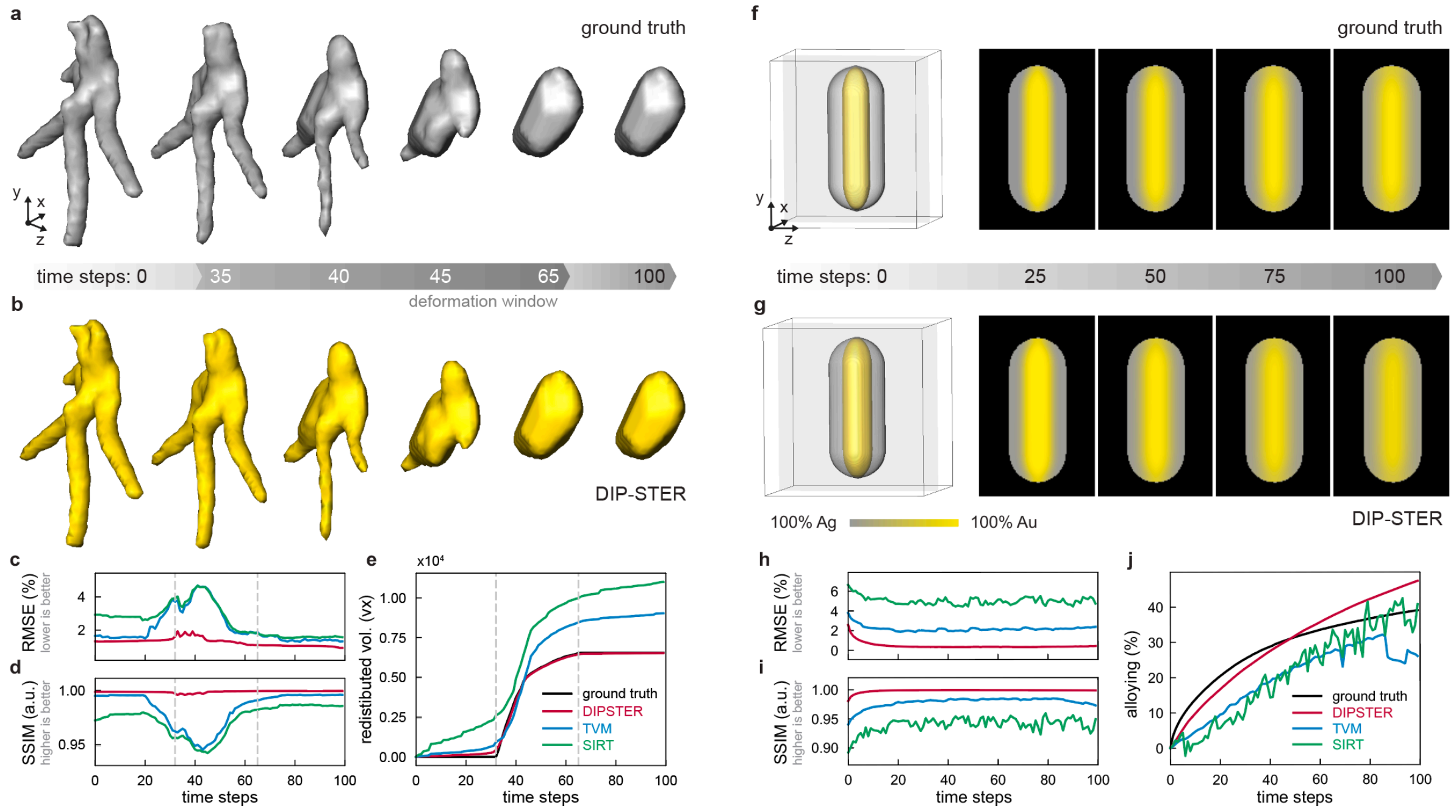

**Fig. 3. Evaluation of DIP-STER reconstructions in ideal acquisition conditions with synthetic Au nanostar and Au@Ag core-shell nanorod**. **a**, Isosurface renderings of the synthetic ground truth nanostar with melting dynamics and, **b**, of the corresponding DIP-STER reconstruction. To introduce non-linear dynamics, the deformation function was only present between time steps 33 and 66. **c**, RMSE and, **d**, SSIM measurements of DIP-STER, TVM, and SIRT reconstructions against the ground truth reference as a function of time. **e**, comparison of the redistributed volume over time. **f**, Isosurface renderings and orthoslices of the synthetic Au@Ag nanorod with alloying dynamics and, **g**, of the corresponding DIP-STER reconstruction. The orthoslices are in the yz plane of the reconstruction frame. **h**, RMSE and, **i**, SSIM measurements of DIP-STER, TVM, and SIRT reconstructions against the ground truth reference as a function of time. **j**, Comparison of the alloying percentage over time.

Reconstructions were then evaluated quantitatively by measuring a range of quality metrics including the structural similarity measure index (SSIM), the root mean squared error (RMSE) and the peak signal-to-noise ratio (PSNR) between the reconstruction and the simulated volume, as well as specific physically-relevant properties, include the surface, volume and redistributed volume for the nanostar, and the alloying percentage for the nanorod. **Fig. 3** shows a subset of these measurements, with complete time-dependent and time-averaged values in **Supplementary Fig. 6** and **7**. For comparison, we also performed similar measurements on conventional SIRT and TVM reconstructions obtained from a moving time-projection window (as illustrated in **Fig. 2a** and **b**, with detailed visual comparison in **Supplementary Fig. 4** and **5**).

Overall, DIP-STER performed strongly on both types of dynamics. High-quality values of RMSE (< 2% of error in either case) and SSIM (> 0.99) were systematically obtained, and at all times. DIP-STER largely outperformed SIRT and TVM reconstructions, both qualitatively as a reduced missing wedge is clearly visible (**Supplementary Fig. 4** and **5**), and quantitatively across all metrics (**Supplementary Fig. 6** and **7**). Although the quality of all reconstruction methods was lowest when the rate of deformation was highest, DIP-STER reconstructions were still highly accurate. For example, the onset of deformation was perfectly recovered (**Supplementary Fig. 8 a** and **b**), and a change as sharp as a complete branch loss between two frames of the nanostar series showed only minor interpolation artefacts (**Supplementary Fig. 9a** and **b**). Thus, in ideal conditions, the temporal resolution of DIP-STER can approach the rate of projection acquisition even on abrupt changes. The early (and fastest) steps of alloying in the nanorod also showed slightly degraded quality metrics, resulting in a core-shell structure with less sharp boundaries than the ground truth and impacting, in turn, the rest of the alloying measurements. Compared to the excellent nanostar results and the precision of the nanorod outer shape (**Supplementary Fig. 5**), this suggests that morphological changes are easier to recover than intensity ones. Still, and in contrast to the other reconstruction methods, DIP-STER measurements clearly retrieved the non-linear dynamics of alloying.

**DIP-STER on realistically corrupted data**

Experimental tilt series are subject to multiple sources of corruption, including Poisson noise, stage drift, scan distortion, imperfect tilt-axis alignment, intensity fluctuations and background signal. These geometry and intensity errors are of particular importance to DIP-STER because they can wrongly be treated as apparent dynamics by the network. As an example, we introduced localized shifts and rotation misalignments in a nanostar tilt series. DIP-STER indeed treated those movements as dynamics, and produced volume series with clear shifts at the affected time steps (**Supplementary Fig. 10**). At the same time, we found that our manifold design can guard against these projection outliers. Producing volumes at the ($t_i$, $y_k$, $\theta_i$) coordinates used during training treated the artefacts as apparent dynamics, whereas volumes at ($t_i$, $y_k$, $\theta = 0°$) coordinates produced a smoother series (**Supplementary Fig. 10**), possibly because those coordinates describe a more compact path on the manifold and more strongly regularize the time dimension.

Next, we tested DIP-STER on a realistic tilt series of the simulated nanostar, with typical experimental corruption added (**Fig. 4a**). We first confirmed that a conventional ET pre-processing for alignment with cross-correlation provided reasonable results. Despite the tilt series not fulfilling the standard assumptions of a static object, shifts were retrieved with a 1-3 pixels accuracy (**Supplementary Fig. 11**). The tilt series was then processed according to standard ET procedures for background removal, and reconstructions were performed with DIP-STER (**Fig. 4b** and **c**, **Supplementary Figs. 12-14**). For this realistic dataset, we added a total-variation (TV) regularization on the loss term (see the **Methods**),[60] which improved noise rejection as further discussed in **Supplementary Note 2**. Results with the ($t_i$, $y_k$, $\theta_i$) coordinates suggest that the processed tilt series still contained residual errors (either of position or intensity) as high frequency fluctuations in the reconstructions were observed (**Supplementary Fig. 12**). In comparison, the 0°-coordinates produced a smoother time series; still captured the non-linear dynamics with high quality (SSIM > 0.98 and RMSE < 4 % at all time points; and produced significantly better results than TVM and SIRT (**Fig. 4c-d** and **Supplementary Fig. 13** and **14**). This confirms that the 0°-manifold is critical in preventing the absorption of high-frequency residual errors and motivates its standard use for DIP-STER inference. We also note that center-of-mass plots show low-frequency positional errors of the volumes, up to 3-4 voxels, even as the physical descriptors were accurate (**Supplementary Fig. 12a**). This is likely linked to remaining errors in the alignment process and will be of limited importance for isolated particles unless the exact position is of interest. Finally, the abrupt onset of deformation was again retrieved with single-time step precision (**Supplementary Fig. 8c**), but the strong inter-frame branch loss was interpolated across 5-6 frames (**Supplementary Fig. 9c**). This suggests that in presence of residual artefacts the spatiotemporal resolution may be reduced and shows that DIP-STER remains bound by the acquisition rate in relation to the speed of deformation and, like conventional tomography, by the quality of the tilt series.

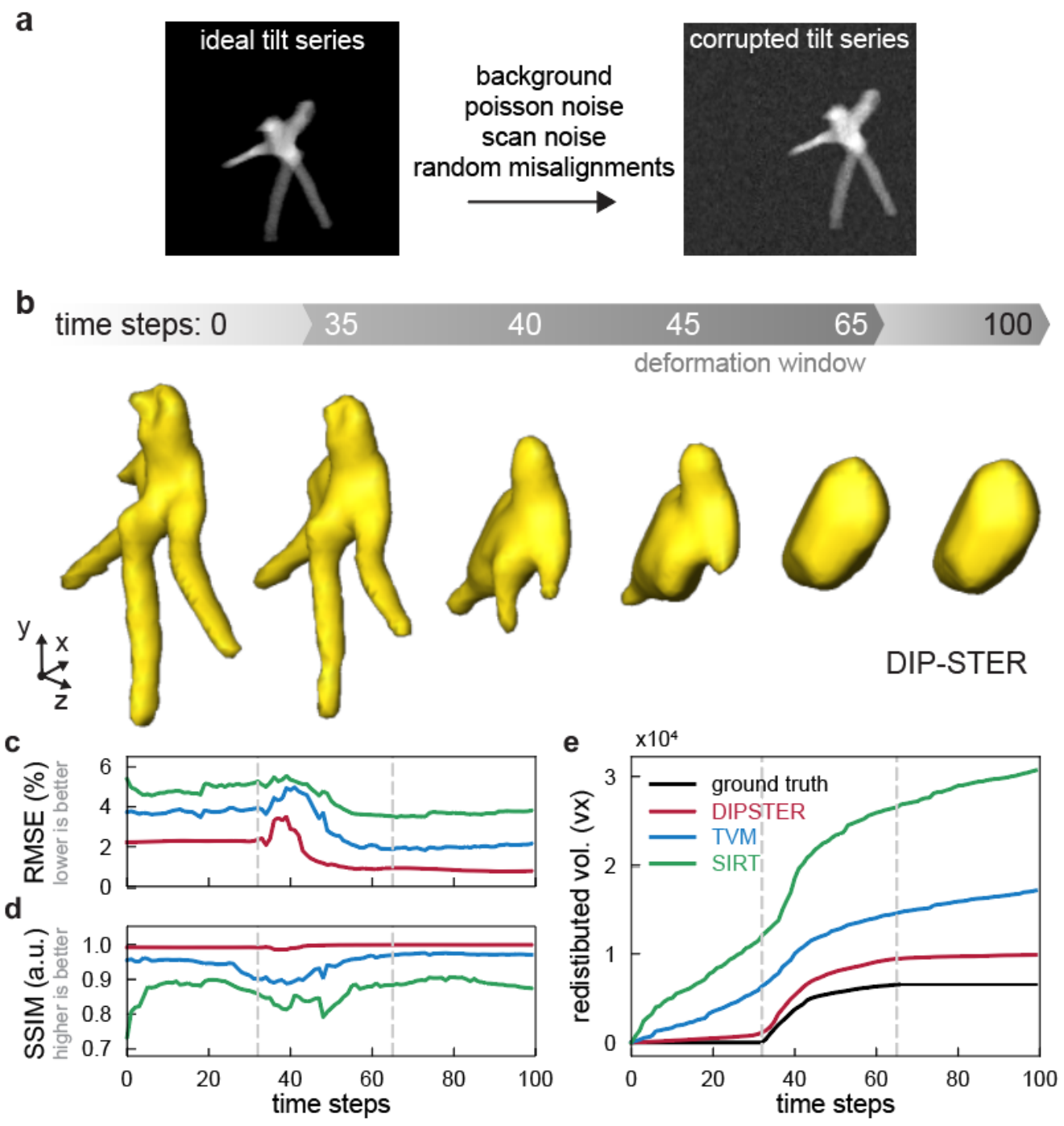

**Fig. 4. Evaluation of DIP-STER reconstructions in corrupted acquisition conditions with synthetic Au nanostar**. **a**, Example projection from an ideal tilt series and after corruption with typical acquisition artefacts. **b**, Isosurface renderings of DIP-STER volumes produced from the pre-processed, corrupted tilt series. **c**, RMSE and, **d**, SSIM measurements of DIP-STER, TVM, and SIRT reconstructions against the ground truth reference as a function of time.

### Tracking nanoscale dynamics in 4D

The merits of the DIP-STER framework were further demonstrated on two experimental samples during applied heating: an Au nanostar and a silica-coated Au@Ag nanocube (**Fig. 5a-c**). These samples were selected because they exhibit unique dynamic behaviour (morphological change and alloying, respectively) in response to heat, which in turn alters optical properties of importance for bio-sensing, therapeutic or catalytic applications.[10,22,23] As such, understanding, tracking and quantifying with fine temporal resolution how they evolve in response to heat is of high importance.[21] The Au nanostar was heated at a constant temperature of 220°C. For the Au@Ag nanocube, heating was first conducted at 300˚C for 12 minutes, followed by a step change to 420°C. These temperatures were chosen to provide significant changes within a reasonable timeframe.[61] As reshaping may occur at such temperature, the silica shell was intended to lock the overall particle shape and isolate the alloying process from other changes.

Experimental acquisitions were conducted during a period of approximately 35 minutes of heating, with tilt series continuously acquired using our modified GRS scheme in the ±70° angular range for the Au nanostar, and ±65° for the nanocube. GRS is an unconventional scheme for ET that requires large angular jumps and backlash correction and makes experiments more challenging than conventional ET. Specifically, it is critical for the eucentric height and scan orientation to be set accurately to minimize shifting and focusing needs during acquisitions and facilitate pre-processing thereafter. We further advise on the practical use of the scheme in **Supplementary Note 3**. In our implementation, we used a script to set the next tilt angle and automatically handle backlash correction while tracking and focusing the region of interest manually. This yielded, on average, 2-3 projections•min$^{-1}$ (**Fig. 5d** and **e**).

A DIP-STER network for each particle was then trained on the acquired tilt series, following the procedure outlined above. After training, the network could compute orthoslices, whose reprojections very accurately correspond to the experimental 2D projection images, both visually (**Supplementary Fig. 15 and 16**) and quantitatively (see the training curves in **Supplementary Fig. 17 and 18**). To produce volume-time series, we again queried manifold coordinates corresponding to all slices of a volume, one at each minute of the experiments. A subset of the series is shown in **Fig. 5a-c**.

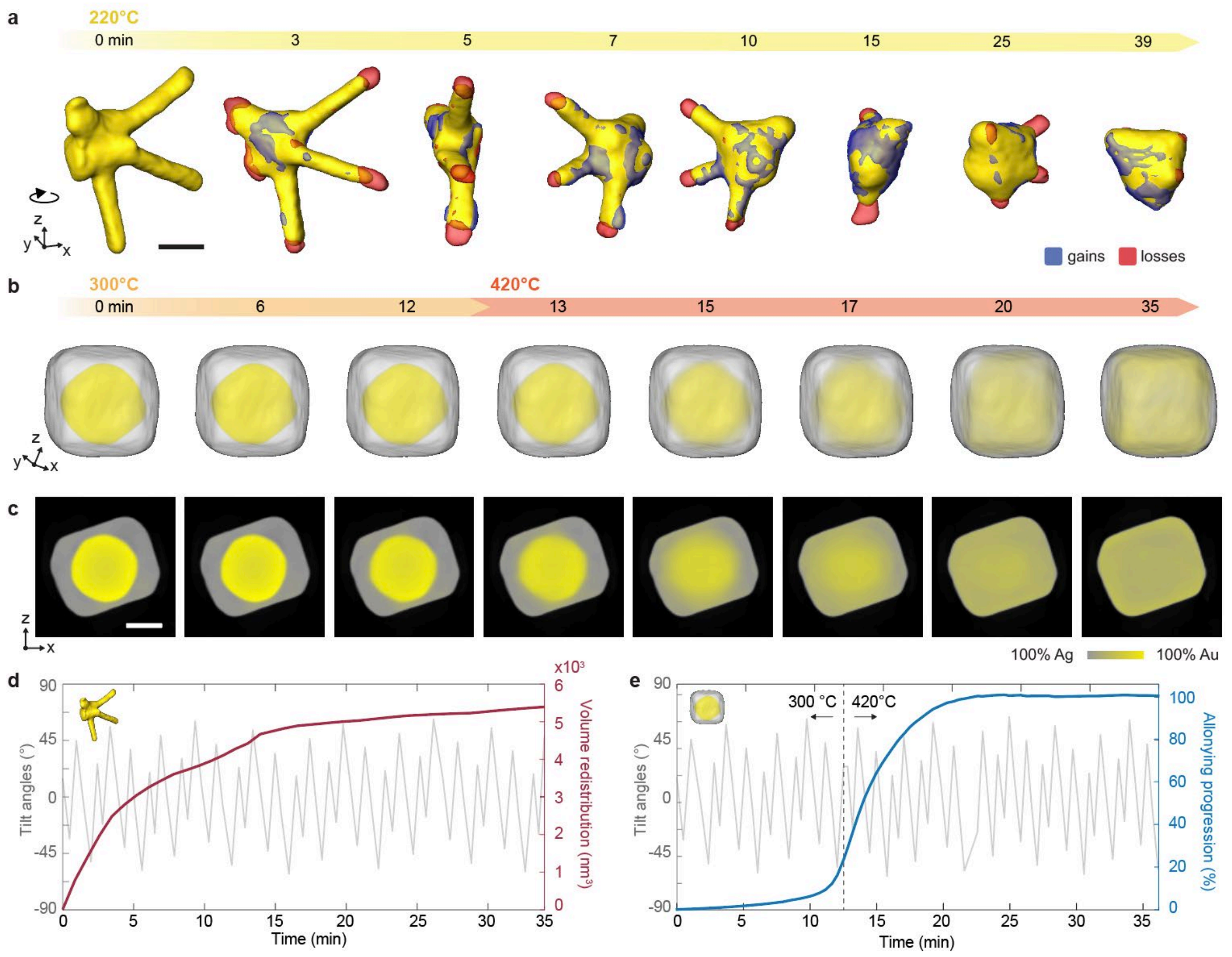

**Fig. 5. DIP-STER reconstructions of experimental, heat-induced, nanoscale dynamics**. **a**, Isosurface renderings at selected time steps of the DIP-STER reconstructions of a nanostar heated to 220°C. Colours indicates area of material loss (red) and gain (blue), between each pairs of displayed time steps, clearly showing loss of branches and redistribution to the core. Scale bar is 20 nm. **b**, Volume rendering and **c**, orthoslice of the DIP-STER reconstructions of a silica-coated Au@Ag core-shell nanocube heated to 300°C then 420°C. The Au core gradually diffuses as Au and Ag alloy. The colour scale is set to map the bimodal intensity distribution to pure Ag and pure Au domains (**Supplementary Fig. 20**).[23] Scale bar is 20 nm. **d**, Experimental projection angles in the nanostar tilt series as a function of time and volume redistribution during the experiment. **e**, Experimental projection angles in the Au@Ag tilt series as a function of time, and alloying progress over time, as estimated from the spread of intensities in the reconstructions.

Both volume-time series evidence the major changes the particles are subjected to during heating. The branches of the nanostar shrunk gradually, while the core grew. To visualize the redistribution process, we measured the difference between the volumes at the displayed time steps and mapped areas of losses and gains (**Fig. 5a**). Assuming constant volume, these difference maps can further be converted to a cumulative redistributed volume and its time-dependent evolution (**Fig. 5d**). Consistent with previous observations,[18] we find that redistribution is faster at early time steps and continuously slows down to a plateau as areas of high curvature shrink. Crucially, the evolution of the particle in this volume series is fully consistent with that of particles not exposed to the electron beam (**Supplementary Fig. 19**) and, despite the relatively low temperature, shows full redistribution of the branches to a near-featureless core by the end of the experiment. This is in contrast to previous reports for similar experiments but using "stop-and-go" ET.[18,26] The prolonged electron beam irradiation during "stop-and-go" ET was hypothesized to degrade the ligand shell into a protective encasing that eventually prevented further reshaping of the specimen.[18,25,27] With DIP-STER, an electron dose at least an order of magnitude lower than the "stop-and-go" approach is needed (**Methods** and **Supplementary Note 4**), which unlocks characterization of nanoscale dynamics previously inaccessible due to the detrimental effects of the electron beam.

The volume-time series of the core-shell nanocube further highlights the capability of DIP-STER to recover fast dynamics not only based on morphological changes but also on quantifying intensity redistribution within the volumes. Throughout the volume-time series, the outer shape of the particle remained stable, confirming the effectiveness of the silica coating. The particle exhibited a sharp core-shell structure at times below the alloying temperature, followed by gradual diffusion and eventually reached a completely homogeneous state (**Fig. 5b**). As with the simulated nanorod, we quantified the alloying progress from the spread of intensities in the histogram of each reconstruction – a lower variance of the intensity distributions indicating higher alloying (**Methods, Supplementary Fig. 20**).[62,63] Measurements confirm the sharp inflection in alloying around the temperature change, followed by a period of progress consistent with diffusion dynamics. Complete alloying was reached within 10 minutes at 420°C, again in full agreement with previous "stop-and-go" ET results.[62,63] Thus, DIP-STER can also serve as the basis for intensity-based quantification of dynamics at a fraction of the experimental time previously required.

## Discussion

Using DIP-STER, we have achieved continuous tracking of nanoscale dynamics in 4D by *in situ* ET. This was made possible by combining a continuous tilting scheme adapted for ET and an implicit network representation of the volume-time series to exploit all the temporally available information. Compared to a moving-window approach with more conventional reconstruction algorithms, the neural architecture of DIP-STER makes better use of the available information and achieves higher fidelity and temporal resolution as shown in simulations. Compared to "stop-and-go" ET as the incumbent approach for (*quasi-*)*in situ* 3D imaging, the combination of DIP-STER and continuous tilting is significantly faster, requiring only a single tilt-time series rather than a full tilt series at each time point. This substantially increases the throughput of dynamic ET experiments while reducing the accumulated electron dose, which is critical to capture the true specimen dynamics. Indeed, the nanostar melting experiments we presented herein revealed dynamics that could not be fully measured hitherto because of beam-induced changes. Moreover, DIP-STER operates continuously, eliminating the need to pause transformations during data collection. This preserves the fidelity of measured transformations, avoiding artefacts introduced by multiple stimulus onsets and pauses, and unlocks the study of transformations that cannot be halted, such as nanoparticle growth.[64]

The implicit spatio-temporal regularization and the training process of DIP-STER are major features of the success of the *in situ* ET experiments that nonetheless entail several challenges. The DL basis of DIP-STER means that the exact nature of the produced volumes, the failure modes and, generally, how a solution is chosen within the possible space are complex questions.

Nonetheless and despite the stochastic nature of network initialization, DIP-STER results are reproducible (**Supplementary Fig. 21**) and well-behaved. We have not observed spurious artefacts or out-of-line results suggesting hallucinations. Evaluation of the spatiotemporal resolution is also not straightforward because DIP-STER operates without a step size as in "stop-and-go" ET or a moving window of precise width, and does not explicitly perform temporal deconvolution. Our simulations show that given ideal conditions, the temporal resolution approaches the rate of projection acquisition, even on abrupt events. Under typical experimental conditions, sharp dynamics can still be recovered but it appears that the (temporal) sampling requirements are more important. Strongly undersampled events may lead to interpolation across several frames. Generally, it is evident that the quality of DIP-STER remains influenced by the number of projections that can be acquired in relation to the rate of deformation. Thus, the size, location, type of dynamics and of *in situ* experiments will affect the results and the local spatiotemporal resolution of DIP-STER-produced volumes. Establishing clear quantification methods is therefore an important topic to explore in future studies. In any case, gains in projection acquisition rate will be beneficial, and could likely be achieved by integrating automated routines for tracking, focusing and eucentric height determination, as well as pre-calibration of the tilting path. We also note that although other continuous tilt schemes such as binary decomposition can afford higher acquisition rates due to smaller angular steps,[15] this does not appear to offer significant advantages over the GRS scheme, probably because the latter better conditions the network to decouple tilting from deformation (**Supplementary Fig. 22**).

Because DIP-STER infers reconstructions through consistency with the projections at specific times, residual geometry and intensity errors commonly found in ET tilt series can be learned as apparent dynamics. This places strong requirements on the quality of the acquisitions and of the preprocessing steps. For example, another experimental tilt series of an alloying Au@Ag nanocube with noticeable detector shadowing at high angles and strong scan distortions produced reconstructions with significant interpolation artefacts (**Supplementary Fig. 23 and 24**). The requirements can be partially lifted by tuning the manifold coordinates during inference to strengthen the temporal regularization and better reject outliers, but this also means that exceptional events within an experiments may be difficult to retrieve. This is generally in line with the priors implied by the manifold and position DIP-STER as better suited for smooth dynamics. To further improve the treatment of real-world datasets, DIP-STER can readily be augmented with a more accurate forward model for joint optimization of the critical preprocessing steps.[45] This augmentation is currently available in DIP-STER (**Supplementary Note 5**), but the information transfer between the network and the model is not evident and is an important direction for future research. Alternatively, the experimental acquisitions could be improved by, e.g., acquiring images with orthogonal scan directions at each tilt step to mitigate scan and drift distortions.[65] If automated, such a step would only involve a minimal time penalty.

Beyond *in situ* experiments, beam-induced damage can also be considered a specific case of dynamic experiment.[66] Our simulations show that DIP-STER is capable of providing an accurate 3D reconstruction at the time step of the first acquired projection. Thus, an undegraded 3D object could be recovered at an exposure equivalent to a single projection, with is promising for the study of highly sensitive specimen, such as MOFs,[67] or halide perovskite nanoparticles.[68]

Finally, the architecture of DIP-STER makes the approach highly modulable. Regularization can readily be added to the loss to include additional prior knowledge about the specimen.[60,69] DIP-STER currently provides explicit regularization in space only, but an explicit and tuneable regularization in the time dimension could help adapting the results to different dynamics. The forward model can also be adapted for other imaging modalities or to include joint alignment and intensity scaling. Furthermore, the manifold geometry can be tuned to encourage learning non-monotonous dynamics, such as periodic behaviour,[46] or to incorporate experimental parameters such as the temperature profile as additional coordinates.

In summary, we used DIP-STER, a self-supervised machine learning architecture, and a GRS tilting scheme to achieve continuous *in situ* ET. The approach was validated on simulated nanomaterials undergoing changes in morphology and in internal density distribution (alloying), and was applied to experimentally acquired metal nanoparticles heated at various temperatures. DIP-STER enables full 3D reconstructions at arbitrary time points from a single, continuous tilt series, unlocking access to volume-time series and quantitative analysis of dynamic nanomaterials properties. Taken together, these features position dynamic ET as a promising, practical and flexible tool for time-resolved 3D characterization across a wide range of electron microscopy experiments.

## Materials and Methods

### Simulations

Two initial $128^3$ $voxels^3$ volumes were used to generate simulated volume-time series: an Au nanostar and an Au@Ag core-shell nanorod. The nanostar starts from the experimental reconstruction volume. The nanorod was simulated as two rods with the same centre and a core/shell intensity set at 1.0/0.3.

To obtain the ground-truth volume-time series, each particle was deformed iteratively for 100 iterations (i.e., simulated time steps). The deformation function for the Au nanostar simulated heating volume redistribution. At each step, a set fraction of voxels were moved from positions with the lowest number of non-zero neighbour voxels to positions with the highest number of non-zero neighbour voxels. This effectively mimics material redistribution from areas of high to low curvature, similar to the general behaviour of melting nanoparticles. This migration only occurs for iterations 33 to 66.

A deformation function for the Au@Ag core-shell nanorod was designed to simulate alloying. At each iteration, the remaining Au core was isolated by masking out all intensities below the pure Au value (set to 1.0 here). Random diffusion from this isolated Au core was simulated by filtering with a Gaussian filter (σ=2.5) followed by five repetitions of a convolution with a randomly generated $3^3$ $voxels^3$ kernel normalized to sum to 1.0. From this diffused core (D), the complete particle was retrieved by enforcing a constant outer shape with masking and a constant total intensity inside the particle, i.e., by recalculating the intensity of every voxel in the updated nanorod (I) using **Eq. 5.**

$$\mathrm{I} = 0.3(1 - \mathrm{D}) + \mathrm{D} \qquad \textbf{(5)}$$

To generate synthetic tilt series that serve as input for DIP-STER reconstruction, a series of 100 angles was generated using the GRS tilt scheme over a range of ±70° (**Eq. 1**). The tilt series were obtained by projecting the volumes from the volume-time series at their corresponding angle, i.e., the volume after deformation step i was projected at the i-th angle. Each of the resulting 2D projections was further converted as 1D projections by splitting the projection along the rotation axis in 1D slices of a single pixel width. The 1D series were used to train DIP-STER networks without further processing.

To include real-world distortions to the synthesized tilt series, misalignments, background intensities, Poisson noise and scan noise were added or applied. Misalignments where randomly generated to shift the projected object within the field of view. For background intensities, a slab was projected at each angle of the tilt series and the resulting intensities added to the corresponding projections. The thickness of the slab was set so that the background-to-object intensity ratio was consistent with that of the data from the Au nanostar experiment. For Poisson noise, the projection intensities were multiplied by a scaling factor, and for each pixel a new value was drawn from a Poisson distribution using the scaled intensity as the number of occurrences. Projections were then scaled back to the original intensity range by dividing with the same scaling factor. The scaling factor was chosen for the background standard deviation to be consistent with that of experimental data under typical acquisition conditions (see next sections). Scan noise was introduced using abTEM with default parameters.[70]

### Synthesis materials

Analytical grade reagents, $HAuCl_4 \cdot 3H2O$ (≥99.9%, CAS: 16961-25-4), silver nitrate ($AgNO_3$; ≥99.9%, CAS: 7761-88-8), l-ascorbic acid (99%, CAS: 50-81-7), Triton X-100 (laboratory grade;

CAS: 9002-93-1), 4-mercaptobenzoic acid (99%, CAS: 1074-36-8), and sodium borohydride ($NaBH_4$, 99%, CAS: 16940-66-2), were purchased from Merck.

All glassware, lids, and magnetic stir bars used for synthesis were cleaned thoroughly with aqua regia (3:1 concentrated hydrochloric acid (HCl; ACS reagent 37%, CAS: 7647-01-0) to nitric acid ($HNO_3$; 68%, CAS: 7697-37-2)) and rinsed thoroughly with MilliQ water three times before use.

**Au nanostar synthesis methods**

The preparation of dendritic Au nanostars was performed based on a published method.[71] This synthesis relies on seed-mediated growth, meaning that it is performed in two sequential steps in two different reaction vials, with the first step involving the formation of small gold nanoparticle nuclei (seeds) which are then overgrown into the final nanostars.

For the seed preparation, 50 µL of 50 mM $HAuCl_4$ was added to 10 mL of 0.15 M Triton X-100 surfactant and mixed well by hand. Then, the solution was placed under vigorous magnetic stirring (> 1000 RPM), and 600 µL of 10 mM ice cold $NaBH_4$ (prepared by adding 5 mL of ice-cold water for every 1.9 mg $NaBH_4$ solid immediately before addition) was rapidly injected. Stirring was maintained for 2 min, then the stirring was stopped and the sample was aged at 4 °C for 10 min.

After seed ageing, seed overgrowth was carried out in a separate vial. First, 0.2 mL of 25 mM $HAuCl_4$ was added to 10 mL of 0.15 M Triton X-100 solution, which was mixed by hand. Then, the solution was brought under vigorous magnetic stirring (>1000 RPM) and 0.08 mL of 10 mM $AgNO_3$ was added, followed by 0.16 mL 100 mM ascorbic acid, and immediately followed by the quick addition of 3.5 µL of the seed solution. Then the stirring speed was decreased to 500 RPM and left to stir for 12 h.

After 12 h, the nanostars were collected by centrifugation at 4000 g and resuspended in the same volume of MilliQ water (10 mL) To stabilize the stars, the Triton X-100 capping ligand was replaced by 4-mercaptobenzoic acid (4-MBA). 10 µL of 1 mM 4-MBA was added slowly to the resuspended Au nanostars under vigorous (> 1000 RPM) stirring. The solution was left for 30 min under the same stirring conditions, then the nanostars were washed *via* centrifugation at 4000 g, redispersed in the same volume of MilliQ water (10 mL), and the 4-MBA addition and incubation was repeated a second time. The stars were finally washed twice *via* centrifugation at 4000 g and resuspended in MilliQ water at the desired volume (~ 1 mL).

**Specimen preparation for *in situ* heating**

Au nanostars were obtained dispersed in water following the method presented above. Au@Ag@$SiO_2$ nanocubes were obtained from a previous work as a dispersion in ethanol.[63] For experimental acquisitions, 3 x 0.5 µL of the sample dispersions were dropcasted and left to dry on MEMS-based heating chips (DENSsolutions Wildfire) previously activated by 2x5" in an 9/1, Ar/$O_2$ plasma (Fischione 1070 nano cleaner, at 30% power, 30 sccm gas flow). For Au nanostars, the chip was cleaned with activated charcoal/ethanol after deposition to reduce residues from the dispersion, as outlined in a prior publication.[72]

**Electron tomography acquisition**

Tilt series were collected using a DENSsolution tomography heating holder in an Osiris microscope (ThermoFischer Scientific, TFS), operated at 200 kV or a TFS Titan cubed operated at 300 kV, both in HAADF-STEM. Prior to acquisition, a beam shower was systematically performed for 15 minutes at 1.5 $e^-/nm^2/s$ to prevent contamination during further imaging. For the Au@Ag nanocubes, the beam shower was conducted at 200°C, which was found to radically mitigate charging and reduce particle mobility in the next steps, while remaining low enough to not induce morphological or alloying changes to the particles.[61,63] For tilt series acquisition, the screen current was set to 60 pA, acquisitions were done at a pixel size of 0.44 nm and 1.6 µs dwell time (nanostar) and 0.27 nm and 1.0 µs dwell time (nanocube). Images of $1024^2$ pixels$^2$ were recorded, resulting in an electron dose of 30 e$^-$/Å$^2$/frame (nanostar) and 51 e$^-$/Å$^2$/frame (nanocube). Tilt series were acquired using GRS with a 2° backlash correction implemented to ensure all angles were approached from the negative side of the alpha-tilting range.[15] The minimum and maximum angles were set to ±70° for the nanostar and ±65° for the nanocube, which were fully accessible with the tomography heating holder. The reduced tilting range for the nanocube was required to prevent detector shadowing at higher angles which reduced the measured intensities. This was not a problem for the nanostar because a physically motivated image normalization could be applied (see next section). Tilting was controlled with TEMScript, whereas focus and tracking adjustments were performed manually (see practical insights in **Supplementary Note 3**). The focus and tracking procedure required on average 20-30 s regardless of the sample, and were performed at lower magnification, typically corresponding to a pixel size of 2 nm. This added an additional ~20 e$^-$/Å$^2$ electron fluence between each acquired projection. In total, the estimated accumulated electron doses were $3.5 \times 10^3$ e$^-$/Å$^2$ for the nanostar, and $5.2 \times 10^3$ e$^-$/Å$^2$ for the nanocube. Heating was applied after the first acquired projection and was set to 220 °C for the nanostar, and to 300 °C, then 420 °C for the nanocube, switching after frame 27, around 12 minutes.

**Experimental tilt series alignment and preparation**

Prior to reconstruction, experimental tilt series were processed for alignment, background removal, and intensity normalization when applicable with in-house codes in Matlab 2024b. For the nanostar, the images were first binned by 2. The median value of each image was subtracted to centre the background intensities to zero, and the background was removed by removing the pixels outside of a mask set at 0.9 times Otsu's threshold. Each image was scaled to the same mean value and to a standard deviation of 1.0 to correct, in a first approximation, for changes in illumination of the HAADF detector. This is motivated by the expectation that melting in this case is purely redistributive. The series was then aligned with cross-correlation and centred within the reconstruction frame using the centre of mass of the averaged tilt series. The orientation of the rotation axis was refined manually by minimizing the typical arc artefacts visible in a conventional ET reconstruction with all acquired projection.[73] Despite the dynamics, arcs were still clearly visible if the rotation axis was shifted or tilted. Finally, the region of interest was cropped and the images were downsampled with bilinear interpolation to be $128^2$ px$^2$. Final voxel size was 0.87 nm.

Pre-processing for the Au@Ag nanocube followed a similar procedure, except for the intensity normalization which was only done by scaling images to the [0,1] range corresponding to the minimum and maximum value of the complete tiltseries. This is because the standard deviation cannot be assumed constant throughout the tiltseries due to the alloying process and motivated the use of a smaller angular range to prevent detector shadowing (**Supplementary Fig. 24**). The final voxel size for the Au@Ag nanocube was 0.73 nm.

**Conventional ET with projection sharing**

To obtain volume-time series with conventional ET reconstruction, a symmetric window was set to include 30 projections around the time for which a volume should be reconstructed. In other words, the reconstruction corresponding to the time at which the i-th projection was acquired used a sub-tilt series with projections i ± 15. For times at the start and end of the complete GRS tilt series, the number of projections in the sub-tilt series was therefore reduced. For example, the reconstruction corresponding to the time at which the 1$^{st}$ projection was acquired used projections 1-15, the reconstruction corresponding to the time at which the 2$^{nd}$ projection was acquired used projections 1-16, etc. For each of these sub-tilt series, a volume was reconstructed with 100 iterations of the SIRT algorithm using the Tomosipo wrapper for the ASTRA toolbox.[74–76] For TVM, the reconstruction was formulated as the inverse problem $\hat{\mathbf{x}} =$ $\arg\min_{\hat{\mathbf{x}}\geq 0} ||H\hat{\mathbf{x}} - \mathbf{m}||_{\ell_2}^2 + \lambda TV(\hat{\mathbf{x}})$where $\hat{\mathbf{x}}$ is the reconstructed volume, **m** the measured data, *H* the projection operator and λ a scalar regularization factor. Optimization was performed using automatic differentiation in PyTorch and the Tomosipo operators for backward and forward projection.[76] The Adam optimizer was used with a learning rate of 0.1 for 150 iterations. The optimal TV weight was selected at the inflection point of the L-curve described by the size of the TV term versus that of the data consistency term.[77,78]

**DIP-STER training and 4D reconstructions**

To obtain volume-time series with DIP-STER, the tomographic reconstruction process was formulated as an optimization problem solving **Eq. 4.** Training was performed with PyTorch using the Geman-McClure loss function implemented with the Kornia library, due to its resilience to noise and outliers. Furthermore, we include a TV regularization term to improve noise rejection from the experimental data.[60] The complete loss is in **Eq. 6.**, where **m** is the measured data or tilt series, **p** is the predicted data (the forward projection of the reconstructed orthoslice), $\mathcal{L}$ (**m, p**) is the loss indicating the difference between the actual and predicted data, TV is the TV term and λ the regularization strength.

$$\mathcal{L}(\mathbf{m}, \boldsymbol{p}) = \frac{2(\mathbf{m}-\boldsymbol{p})^2}{(\mathbf{m}-\boldsymbol{p})^4+4} + \lambda \mathrm{V}(\boldsymbol{p}) \quad \textbf{(6)}$$

Given the loss function in **Eq. 6.**, the full training procedure can be described (for the λ = 0 case) with **Eq. 7.**

$$\{\boldsymbol{\varphi}^*, \boldsymbol{\psi}^*\} = \mathrm{argmin}_{\boldsymbol{\varphi},\boldsymbol{\psi}} \frac{1}{\mathrm{N}} \sum_{\mathrm{i}=0}^{\mathrm{N}} \sum_{\mathrm{k}=0}^{\mathrm{K}} \frac{2\left(\mathbf{m_{i,k}} - \mathrm{H}\left(\mathrm{f}_{\boldsymbol{\varphi}} \circ \mathrm{g}_{\boldsymbol{\psi}}\left(\mathrm{c_i}(\theta_\mathrm{i}, \mathrm{t_i}, \mathrm{y_k})\right)\right)\right)^2}{\left(\mathbf{m_i}, \mathbf{k} - \mathrm{H}\left(\mathrm{f}_{\boldsymbol{\varphi}} \circ \mathrm{g}_{\boldsymbol{\psi}}\left(\mathrm{c_i}(\theta_\mathrm{i}, \mathrm{t_i}, \mathrm{y_k})\right)\right)\right)^4 + 4} \quad \textbf{(7)}$$

At the learning stage, DIP-STER training is performed using random i and k coordinates from 85-95% of the measurements, for 20-50 epochs. The 5-10% held-out frames are used for validation, and enable monitoring overfitting to experimental noise and hyperparameter optimization. Important hyperparameters include the MAPNET and CNN size, batch size, number of training epochs, learning rate, and regularization weight. We have found these parameters to be generally well-behaved and easy to tune on the simpler cases such as single-material samples (**Supplementary Note 2** provides guidelines for optimization). For more challenging specimen, automated Bayesian tuning with Optuna is included.[79] Back propagation was implemented with an Adam Optimizer using gradient descent, with the SSIM and loss monitored every 100 iterations using WanDB. Training typically required < 4 Go of VRAM and ran within 0.5-2 hours on computers supporting a Nvidia RTX4090 GPU with 24 GB of memory or a Nvidia RTX5090 GPU with 32 GB of memory.

To reconstruct volume-time series, a list of time steps and y-positions are passed to a trained DIP-STER network and the predicted time- and y-position-dependent orthoslices are restacked as a 4D dataset.

**Quantification of the reconstruction quality**

After convergence of the DIP-STER networks, reconstruction quality was evaluated with several quantification metrics. For simulated samples, the reconstructed volumes were compared to the simulation at the same acquisition time. For experimental samples, the reconstructions were forward projected to match the acquisition time and angle of the acquired projections. To avoid background signal contributing to the measurement, reconstructed samples and tilt series were masked using Otsu's method prior to measurement.

The RMSE measures the difference between an acquired image (x) and a reference image (y) as a percentage indicating the difference between the two images as a pixel-by-pixel comparison as the average of all pixels ($N_p$) (**Eq. 8)**.

$$\mathrm{RMSE}(x, y) = \sqrt{\frac{1}{N_p}\sum_{k=0}^{N_p}(x_k - y_k)^2} * 100 \quad \textbf{(8)}$$

The SSIM is defined by **Eq. 10.**, where μ is the average intensity, σ is the variance of the image, and $\sigma_{xy}$ is the covariance of the image. The C values are constants specified in Scikit-Image.

$$\mathrm{SSIM}(x, y) = \frac{(2\mu_x\mu_y + C_1)(2\sigma_{xy} + C_2)}{(\mu_x^2 + \mu_y^2 + C_1)(\sigma_x^2 + \sigma_y^2 + C_2)} \quad \textbf{(9)}$$

Both metrics were calculated using their Scikit-Image implementations.[80]

**Measurements of physical properties**

Measurements of the physical properties included the surface, volume, volume redistribution, and alloying progress.

*Morphological metrics (nanostar volume-time series)*. The reconstructions were binarized with Otsu's threshold. The volume was the number of non-zero voxels after thresholding. The number of redistributed voxels was obtained by subtracting temporally adjacent volumes and converting to a volume. For the experimental nanostar, we considered 0.87 nm voxel size to convert to redistributed volume. The surface was calculated as the number of voxels remaining after subtracting the binarized reconstructions to a volume subjected to a single binary dilation step.

*Alloying progression (Au@Ag nanorod and nanocube volume-time series)*. The alloying percentage calculation followed the procedure originally presented in ref.[23] Briefly, the principle relies on converting the spread of intensities (standard deviation of the intensity histograms) to the alloying progress following:

$$a_i = \frac{\sigma_i - \sigma_0}{\sigma_\infty - \sigma_0} \qquad \textbf{(10)}$$

With $a_i$ and $\sigma_i$ the alloying degree of alloying and the standard deviation of the intensity histogram at step i, respectively. $\sigma_0$ and $\sigma_\infty$ are the standard deviations at initial core-shell and perfectly homogeneous particles. All measurements were done within a mask obtained by Otsu's threshold and a single binary dilation step on the simulated volumes, to exclude contributions from the background. It is apparent that when fully alloyed, $\sigma_\infty$ only comes from the imperfect reconstruction process. For SIRT and TVM moving-window series, this effect was estimated following the procedure outlined in ref.[62]. Briefly, the last reconstructed volume was binarized at Otsu's threshold, then multiplied by a scaling factor equal to the mean of the reconstructed interior intensities at the last time step. 15 projections following the GRS tilt scheme were obtained from these ideal volumes. SIRT and TVM reconstructions were computed using similar parameters as outlined above, and $\sigma_\infty$ was measured on these final volumes. Estimating $\sigma_\infty$ is more complex for a non-deterministic approaches like DIP-STER. In a first approximation, the process was considered ideal and, therefore, $\sigma_\infty$ was set to be null for the simulations. For the experimental Au@Ag results, the extended final plateau in $\sigma_i$ made apparent that full alloying was achieved and $\sigma_\infty$ was set to be equal to $\sigma_i$ at the last time step.

## Author contributions

T.M.C., A.K. and S.B. conceptualized the DIP-STER framework; T.M.C., R.G., and A.M. developed the DIP-STER code and conducted experiments and data analysis; T.M.C., R.G. and S.B. designed the experiments; G.A.V.-W. conducted the nanostar synthesis; L.L.M. supervised the synthesis; R.G., T.M.C., and A.M. wrote the manuscript and made figures; R.G. and S.B. revised the manuscript and supervised the research; All authors contributed to the discussion of results and reviewed the manuscript.

## Acknowledgements

Ana Sánchez-Iglesias at CIC biomaGUNE is gratefully acknowledged for providing the Au@Ag nanocubes. T.M.C. acknowledges support through funding received from the European Union's Horizon Research and Innovation Program under grant agreement no. 860942 (HEATNMOF). R.G. acknowledges the support of a FWO fellowship under award 12A1V25N. G.A.V.-W. acknowledges support from the European Union's Horizon Europe Research and Innovation Program under the Marie Skłodowska-Curie grant agreement no. 101105300 (PLASMOSTEMFATE). The authors acknowledge financial support by the European Union (ERC SyG No. 101166855 CHIRAL-PRO to L.L.M. and S.B.).

## Declaration of competing interest

The Authors declare that they have no conflict of interest.

## Data availability

The tilt series, pre-trained models and reconstructions used in this publication will be made available in a Zenodo repository.

## Code availability

The code to use DIP-STER and to reproduce this analysis is openly available on a GitHub repository (https://github.com/EMAT-tomography/dipster). A Code Ocean capsule with detailed notebooks will be made available upon publication.

# Supplementary Information

## Supplementary Note 1: Computational Footprint

The process of reconstructing dynamic 3D data over time presents a significant computational challenge. The primary difficulty is that these datasets are subject to the "curse of dimensionality", where the expansion of the dataset comes with a significant computational overhead. For instance, adding an additional time point to a video consisting of 2D images requires adding an additional image to the video or $n^2$ pixels; where n is the number of pixels. To achieve the same result for a volume-time series, one would need to add a full volume or $n^3$ voxels to the dataset. Hence, a modest data set comprising a $512^3$ voxel volume over 100 time points, represented as floating-point numbers, demands approximately 107 GB of memory. This volume exceeds the memory capacity available for graphics processing units (GPUs) in most commercially available desktop computers. Our approach addresses this challenge by computing the time and depth as functions instead of handling the full data set directly. This allows for a more manageable architecture comprising of 2D and 1D network layers. The advantage is a significantly compressed neural network representation of the data, reducing the memory footprint dramatically. Moreover, the scaling of data is efficient: increasing the volume of a standard reconstruction linearly (by '*n*' voxels in every direction) results in a cubic ($n^3$) increase in data requirements but only a quadratic ($n^2$) increase in the neural network size. Whereas reconstructions were performed for projections of 128 pixels$^2$, the network can easily be expanded to support reconstructions of 512 pixel$^2$ projections when following the tuning instructions outlined in **Supplementary Note 3.**

Additionally, as the reconstructed volume-time series is computed from the trained neural network, the neural network can be stored in place of the full volume-time series for compressed on-disk storage. When analysis of the volume-time series is necessary, the desired orthoslices or volumes can be calculated on demand from the neural representation. For the nanoshell volumes-time series (128 voxels$^3$, 100 times, float) the theoretical data size is 1.67 GB, which is stored on a network of 16.9 MB. Likewise, a dataset of 512 voxels$^3$, 100 times, float (107 GB) would be stored as a network of 258 MB.

## Supplementary Note 2: Reconstruction Optimization

In the DIP-STER framework, multiple parameters must be tuned in order to optimize the reconstructed image quality. In this section, we provide a minimal intuition for some of the most critical parameters and typical values used for the reconstructions herein (see also **Supplementary Table 1** for the hyperparameters used for each reconstructions herein). Optimization can nonetheless be challenging, and DIP-STER therefore integrates automated hyperparameter tuning with Optuna.[79]

*Training set and epochs*: The most important parameters to evaluate for deep learning training are the amount of training data and number of training epochs (during an *epoch*, the network will see the entire training set once). Finding the right balance in these hyperparameters is key to obtain exploitable results. With too little epochs, the model will not have converged to a good solution yet. At the same time, DIP networks benefit from an early stop before they start overfitting noise. Based on our tilt series of 70 to 100 projections, training between 25 to 100 epochs was a good range with overfitting appearing around 120 epochs.

*Batch size*: The batch size is the amount of data at each training step (within a *step* the losses are accumulated and the neural network weights are updated at the end). In general, a larger batch size speeds up training at the cost of a larger memory usage but also influences convergence as the network weights are less often updated. In DIP-STER, we typically saw that increasing the batch size above 4 made the convergence of the model slower. We hypothesize that increasing the batch size tends to average the changes needed for the weights too much for an efficient convergence. This would imply that there are big differences in the way the model treats the information for data points that are far in the manifold.

*Network depth and width*: The number of layers in the mapping network (MAPNET) and convolutional neural network (CNN) represents the network depth. The number of fully-connected neurons in each layers of the MAPNET sets the network width. Intuitively, the CNN controls the resolution whereas the mapping layer depth controls the expressiveness of the network e.g. the complexity of temporal & spatial correlations. In the synthetic alloying nanorod for example, DIP-STER needed to capture more complex spatial dynamics due to the larger intensity distribution. As a result, increasing the mapping network depth yielded better results after convergence. **Supplementary Figure 2** shows our baseline architecture. Typically, two fully connected layers of 512 neurons in the MAPNET and two convolutional layers between each up sampling of the CNN were used.

*Learning and decay rate*: The learning rate (α) determines the amount the network weights and biases are updated after each training step. The DIP-STER network uses an ADAM optimizer and StepLR scheduler common to DIP networks. The Adam optimizer modifies the local learning rate per training step *i* based on noise and the loss gradient. The StepLR scheduler reduces the global learning rate by a decay coefficient (γ) every set amount of training steps (s) (**Supplementary Eq. 2**)

$$\alpha_i = \alpha_0 \gamma^{\left\lfloor \frac{i}{s} \right\rfloor} \quad (1)$$

The purpose of a decaying the learning rate is to allow rapid changes during initial learning phases to avoid being trapped in a sub-optimal local minimum, while reducing the rate of change in later learning stages to allow the network to converge.

We have found an initial learning rate at $1 \times 10^{-4}$ to be typically adequate for DIP-STER, while two configurations for the decay rate and number of steps were investigated. The first one is based on a rapidly, but small decay rate steps (γ = 0.985 and $s$ = 32 x total number of steps), while the second relies on bigger steps but at a varying at a slower rate (γ = 0.85 and $s$ = 16 x total number of steps). The first configuration gave faster convergence for the simulated reconstruction, while the second was better for the experimental reconstruction. In practice, this difference is in the order of one to two epochs, which makes these configurations an optimal range for DIP-STER in general.

*Regularization*: DIP networks can also be regularized to improve their behavior in the presence of noisy training data.[60,69,81] Generally, even non-regularized DIP network are effective against noise because their convolutional nature promotes learning high-level signals before those with high impedance like noise.[53] However, this implies that noise rejection comes from underfitting the training data. In the case of DIP-STER, this approach can create conflicts between the need to accurately learn the reconstruction process as well as the spatio-temporal dynamics, and that of not over-training to maintain the noise rejection capabilities. To better handle real-world data, we therefore added a regularization term to the training loss in the form of a total variation (TV) operator on the DIP-STER reprojections.[60] Effectively, the TV operator promotes locally constant intensities by favoring sparsity in the image gradients. We control the amount of regularization against data fidelity with the weight $\lambda$ so that the total loss was (**Supplementary Eq. 3**):

$$\mathcal{L}(\mathbf{m}, \boldsymbol{p}) = \frac{2(\mathbf{m}-\boldsymbol{p})^2}{(\mathbf{m}-\boldsymbol{p})^4+4} + \lambda \mathrm{V}(\mathbf{p}) \quad (2)$$

Where $m$ is the measurement data, $p$ the prediction from DIP-STER forward projected and $V(\boldsymbol{p}) = \sum_n |p_{n+1} - p_n|$ is the 1D TV of $p$ with $p_n$ the pixel value at position *n*. In our case, we found

that a regularization weight between 5.10$^{-6}$ and 1.10$^{-5}$ allowed for good reconstruction fidelity without overfitting noise.

## Supplementary Note 3: GRS implementation and alternative continuous schemes

To perform GRS acquisition at the microscope, the sample is tilted following **Supplementary Eq. 1** (see also **Supplementary Fig.1**).

$$\theta_i = i(\theta_{max} - \theta_{min})\frac{1+\sqrt{5}}{2}\left(\mathrm{mod}(\theta_{max} - \theta_{min})\right) + \theta_{min} \quad \textbf{(3)}$$

We handle the tilting process using TEMscript,[82] so that the next tilt angle in the GRS sequence is automatically set using a simple key stroke. The script also automates backlash correction so that when moving from high to low angles, the stage is overtilted by 2° before reaching to the desired angle. This way, all tilt angles for acquisition are approached from the lower-to-higher angle direction. The script is available as part of the **Supplementary Materials**.

Because of the large tilt angles imposed by the GRS formula, the specimen will experience significant shifts and defocus upon tilting. To maximize the acquisition rate, it is critical that the optical axis aligns as best as possible with the rotation axis of the specimen holder and that the eucentric height is precisely tuned. Importantly, setting or changing the conditions of an *in situ* experiment may change the position of the specimen and move it away from eucentric height. For example, the heating experiments herein typically involved a 2-4 μm change in eucentric height, possibly due to bulging, of the DENS nano-chip. This shift was reproducible and near linear with temperature in the range we used, and could be predicted either through a pre-measurement at a benign temperature, or from previous experiments. Thus, the eucentric height could be finely tuned prior to the experiment, before simply moving to the predicted z-height once at the desired temperature without going through the tuning routine again. Doing so minimize the setup phase time to start acquisitions as quickly as possible. Although some amount of shift and defocus will remain, it is noteworthy that their direction is usually predictable based on the tilt direction (towards higher or lower angles). With these insights, a trained operator can reach down to ~20-25 s per projection acquisition routinely.

It is noteworthy that GRS is not the only tilting scheme for continuous tomography, and an alternative could be the binary decomposition (BD) schemes we have previously adapted for for the electron tomography geometry.[15,83] A BD scheme also go back and forth between the limiting angles but linearly subdivides the total angular range (see an example in **Supplementary Fig. 22**). This results in larger steps (typically 16.5° in our implementation) than a typical incremental scheme (typically 2-3°), but smaller ones that GRS (in which steps up to 80° can be used). In our experience, the BD scheme allows about 2x faster focusing and tracking as a result (10-15s on average against 20-25s for GRS). Yet, in simulations, a DIP-STER model trained with the BD scheme performed only as well as a GRS-based model trained with half the projections. (**Supplementary Fig. 22**) This supports the intuition that large steps are expected to better condition the network to decouple true dynamic changes in the projections from a changes in view angle. It is nonetheless possible that certain dynamics with linear or periodic characteristics may benefit from BD, and that BD may prove easier to fully automatize with more predictable shift-defocus changes between angles. In any case, the DIP-STER architecture is flexible and can serve as the basis for a more systematic optimization of the angular sampling conditions.

## Supplementary Note 4: Electron Dose for *in situ* Electron Tomography

In comparing DIP-STER with "stop-and-go" approaches to (*quasi-*)*in situ* ET, the accumulated electron dose stands out as a major difference. For example in ref.[18], Vanrompay *et al.* performed 10 consecutive ET acquisition covering a total 1200 s of cumulated heating; in ref.[62], Skorikov *et al.* performed 11 acquisitions covering 300 s of heating; and in ref.[63], Mychinko *et al.* covered 600 s of heating with 10 acquisitions. These acquisitions were performed in the so-called "fast ET" approach, in which typically 350-400 images are acquired within 5-6 minutes of continuous acquisition.[18] With otherwise equivalent imaging conditions as used for DIP-STER acquisitions (see main **Methods** – $1024^2$ $px^2$ images, 60 pA beam current, 0.44 nm pixel size), this would amount, as a lower bound estimate, to 6 x $10^4$ $e^-/Å^2$ of accumulated dose over the course of an experiment. In contrast, the electron doses for the DIP-STER acquisitions in this manuscript were 3.5x$10^3$ $e^-/Å^2$ for the nanostar, and 5.7 x$10^3$ $e^-/Å^2$ for the Au@Ag nanocube (see main **Methods** for complete details). Thus, besides enabling continuous 3D imaging for *in situ* ET, DIP-STER allows to cover longer experiments with drastically reduced exposure to the e-beam compared to the stop-and-go approach.

## Supplementary Note 5: Joint Alignment with DIP-STER

Experimental tilt series are susceptible to sample drift, rotation, or scan distortions. To compensate for these artefacts without requiring additional preprocessing, we interlace an alignment optimization routine into the training of DIP-STER using the automatic differentiation framework of PyTorch. After the backpropagation step at each epoch, we optimize, independently and for each input image, a planar affine transformation.

$$\begin{pmatrix} x \\ y \end{pmatrix} = \begin{pmatrix} a & b \\ c & d \end{pmatrix} \begin{pmatrix} X \\ Y \end{pmatrix} + \begin{pmatrix} e \\ f \end{pmatrix}$$

which displaces every pixel coordinate $(X, Y)$ of the reference projection to a corrected position $(x, y)$ using six parameters $(a, b, \ldots, f)$ whose combinations determines the shift, rotation, shear and scale applied to the image. These parameters are optimized by minimizing the following loss function.

$$\mathcal{L}_{\text{aff}} = \left(1 - \text{SSIM}\left(\text{A}(\mathbf{m_{i,k}}), \text{H}\left(\text{f}_{\boldsymbol{\varphi}} \circ g_{\boldsymbol{\psi}}(\mathbf{c_{i,k}})\right)\right)\right)$$

where $\text{A}(\mathbf{m_{i,k}})$ is the affine-transformed measured projection, $\text{H}\left(\text{f}_{\boldsymbol{\varphi}} \circ g_{\boldsymbol{\psi}}(\mathbf{c_{i,k}})\right)$is a 2D projection of a complete volume estimate by DIP-STER.

As DIP-STER optimizes the volume reconstruction to be continuous over time, the goal of this alignment method is to absorb the residual differences between the input projection and the reprojection of the volumes. In practice, we did not find the transform to provide significant improvements beyond a simple cross-correlation alignment (**Supplementary Fig. 11**).

## Supplementary Table 1: hyperparameters of DIP-STER networks

| Sample | Figure | Epoch s | LR | TV | MAPNET depth | MAPNET width | Batch size |
|---|---|---|---|---|---|---|---|
| NS - no distortions | 3b-e | 44 | 1.00E-04 | 1.00E-07 | 2 | 512 | 4 |
| NR - no distortions | 3g-j | 62 | 1.00E-04 | 1.00E-08 | 2 | 512 | 4 |
| NS - with distortions | 4b-e | 58 | 1.00E-04 | 1.00E-05 | 2 | 512 | 4 |
| Au NS - experimental | 5a,d | 50 | 1.00E-04 | 5.00E-06 | 2 | 128 | 4 |
| Au@Ag NC - experimental | 5b,c,e | 50 | 1.00E-04 | 5.00E-06 | 3 | 256 | 4 |

**Supplementary Table 1. Hyperparameters of DIP-STER networks**. NS: nanostar, NR: nanorod, NC: nanocube

## Supplementary Fig. 1-24

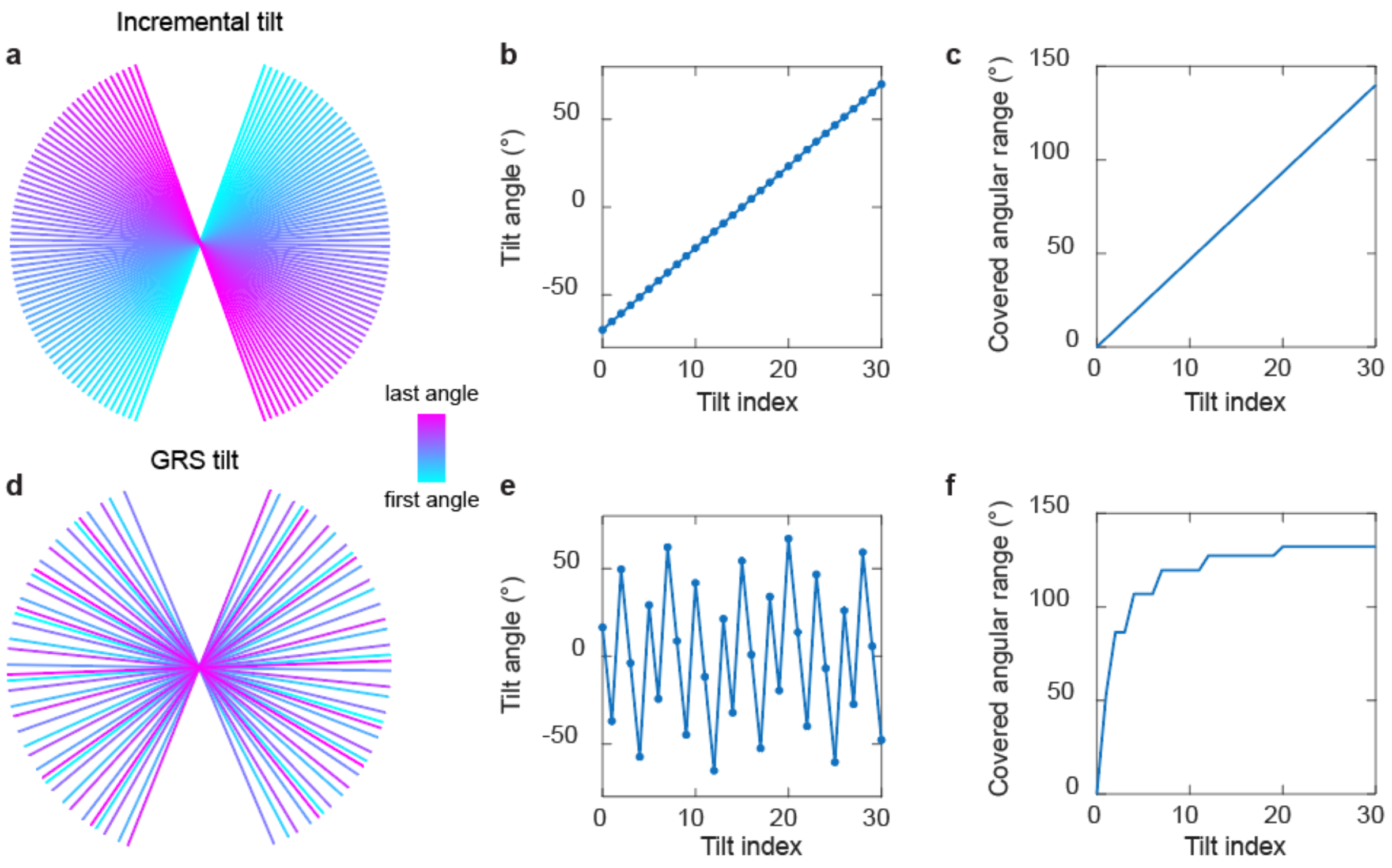

**Supplementary Fig. 1. Comparison of continuous and golden ratio scanning (GRS) tilt schemes** in the [-70°,70°] angular range. (a, d) Visualization of the tilting scheme, (b, e) tilt angles and (c, f) covered angular range for a tilt series of 30 images with the continuous tilt scheme or the GRS tilt scheme as implemented in ref.[15] The GRS scheme rapidly covers the majority of the maximal angular range accessible to the holder.

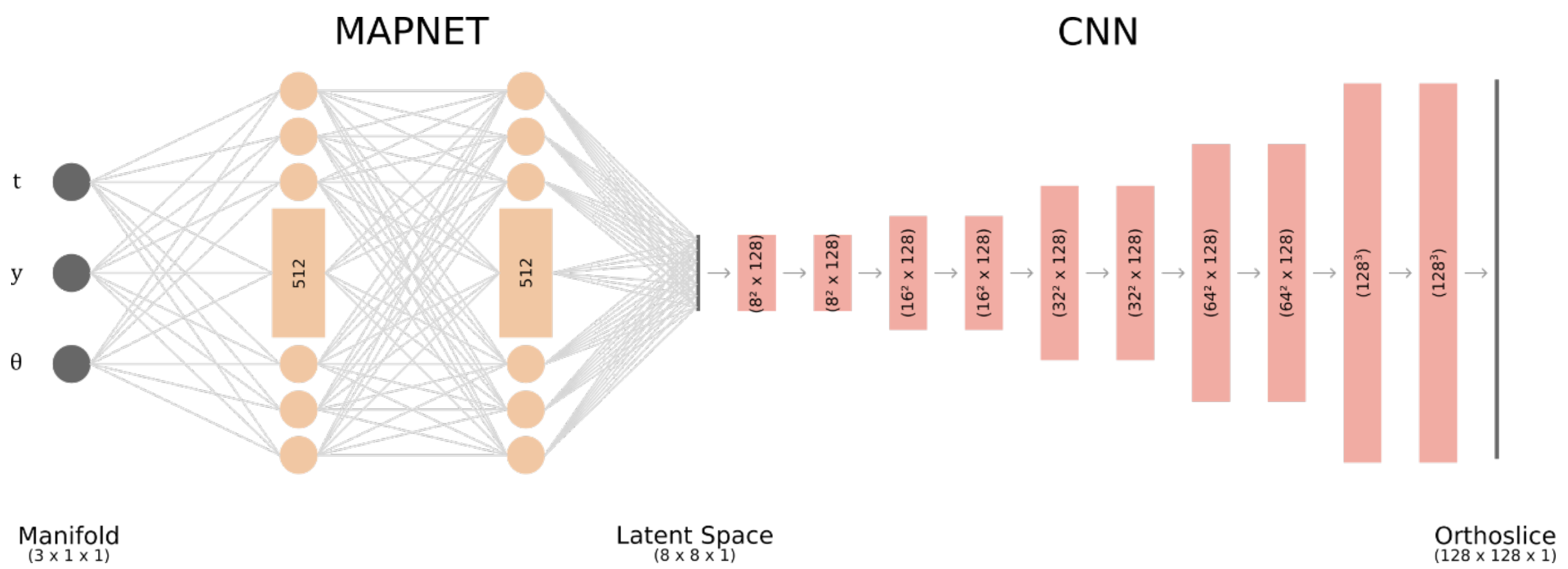

**Supplementary Fig. 2. Schematic of the neural network architecture**. It is composed of two parts, the MAPNET that takes the three coordinates describing the random selected projection and gives a compressed latent vector, and the generative CNN that interpolate the 2D orthoslice associated with the MAPNET latent vector.

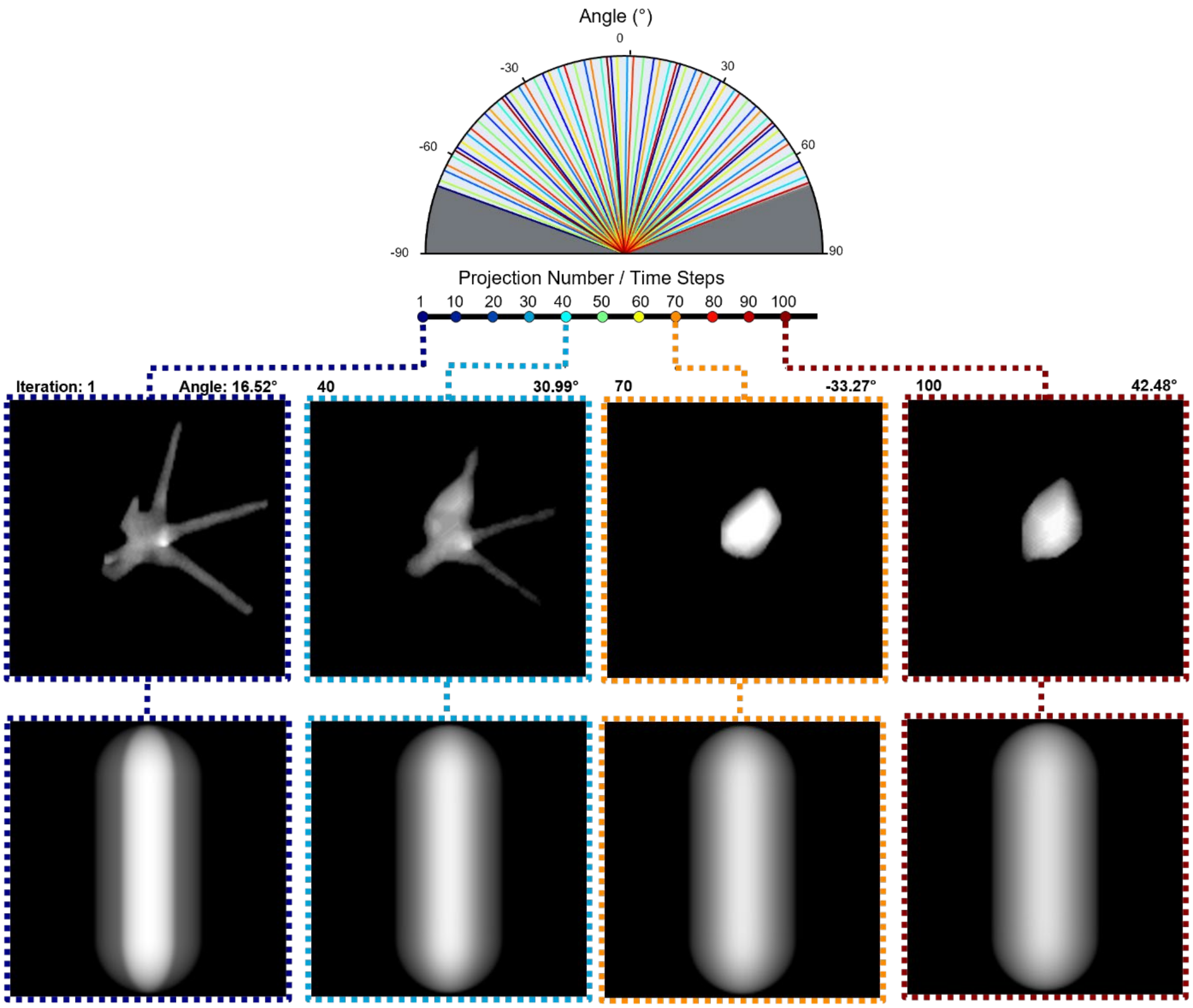

**Supplementary Fig. 3. Simulated tilt series with the GRS tilt scheme**. For each angle in the tilt scheme, a projection (bottom) was obtained by forward projecting a volume (top) in the volume-time series with the angle obtained in the same sequential order in the tilt series. In other words, the first projection was obtained by forward projection a volume after 1 iteration of deformation with the first angle in the tilt series (θ=-70°) obtained in chronological order. This method was used to obtain a tilt series that progressively deforms as projections are acquired for both a nanoshell undergoing simulated beam damage and a nanorod undergoing alloying induced deformation. For each case study, selected volumes and projections are shown.

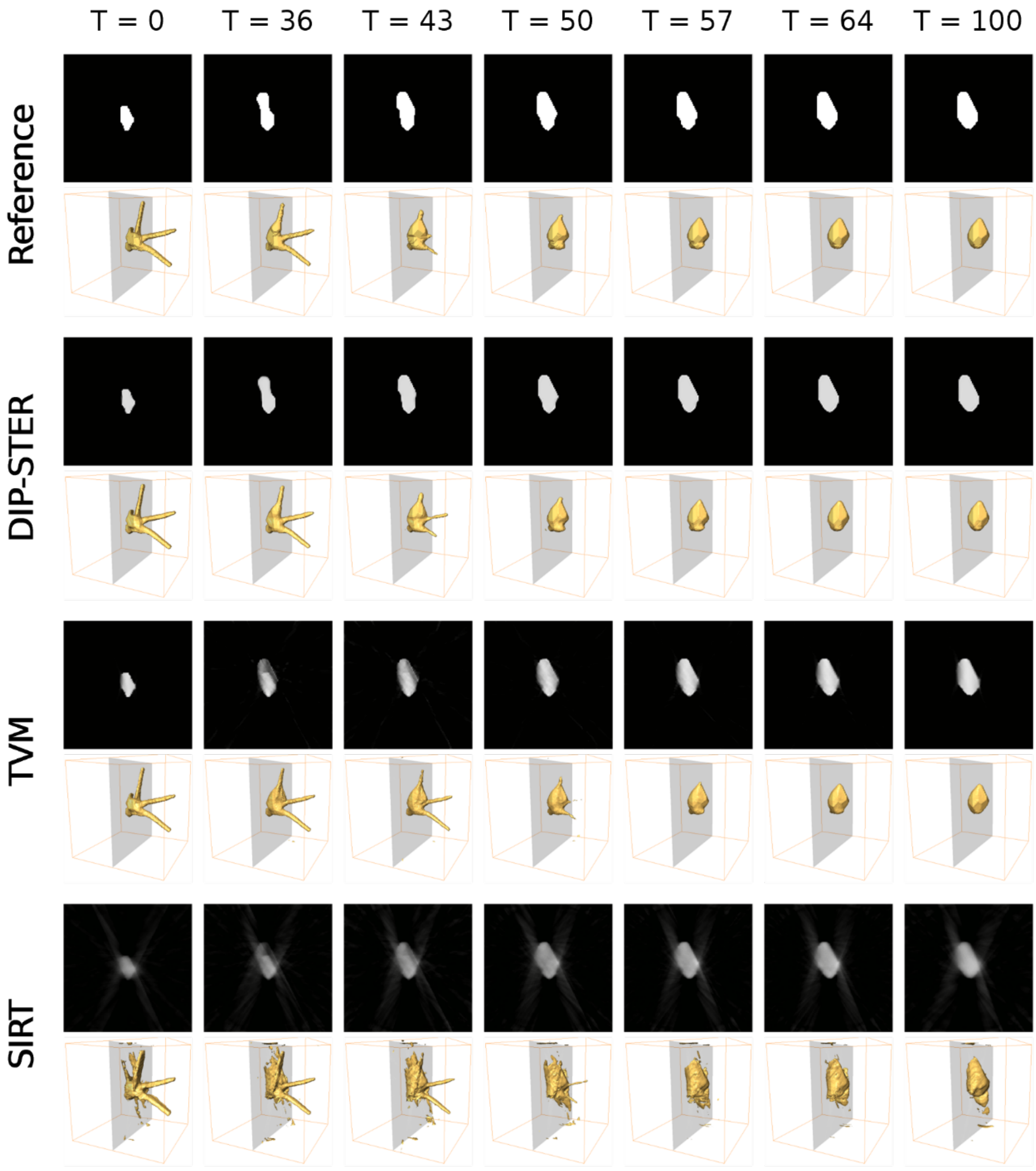

**Supplementary Fig. 4. Comparison of the reconstruction of a simulated nanostar** with redistribution of the volume while heating with DIP-ST ER, TVM and SIRT. The reconstruction using DIP-STER is in close agreement with the simulated ground truth. In particular, we can accurately identify the shrinking of the branches reconstructed structures. SIRT reconstructions suffer from stronger missing wedge artefacts caused by the low amount of projections available in the time windows selected. TVM does not have noticeable defects in the orthoslice but the size of the branches during the deformation is not accurate. We can notice progressive redistribution in the overall shape of this nanostar.

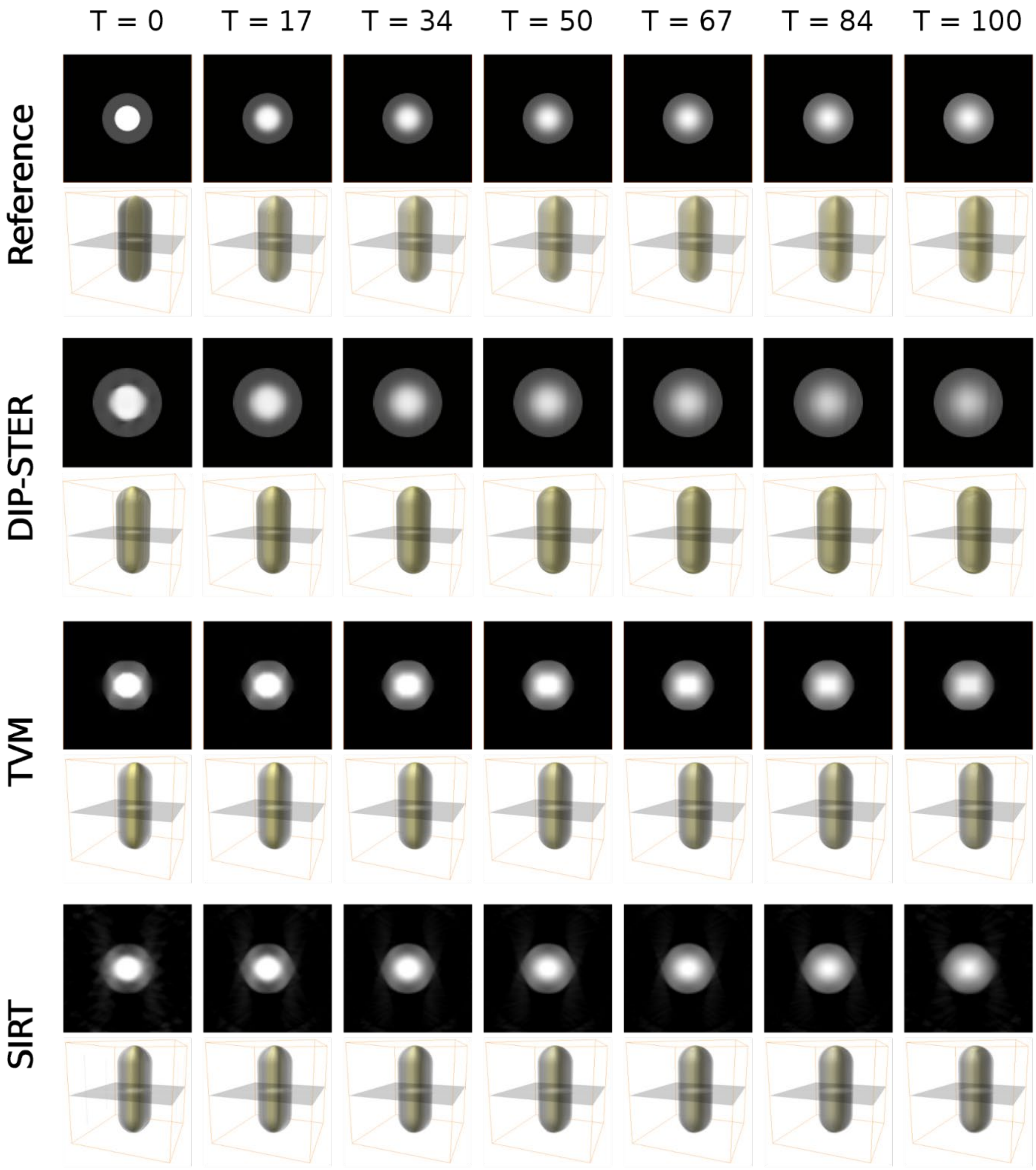

**Supplementary Fig. 5. Comparison of the reconstruction of a simulated nanorod** during alloying. Here again the reconstruction with DIP-STER is more precise than with SIRT. The isosurface of the pure Au core clearly shrinks in the DIP-STER reconstructions, in line with the alloying progression. On the other hand, the core isosurface in the SIRT reconstructions show limited changes across the volume-time series.

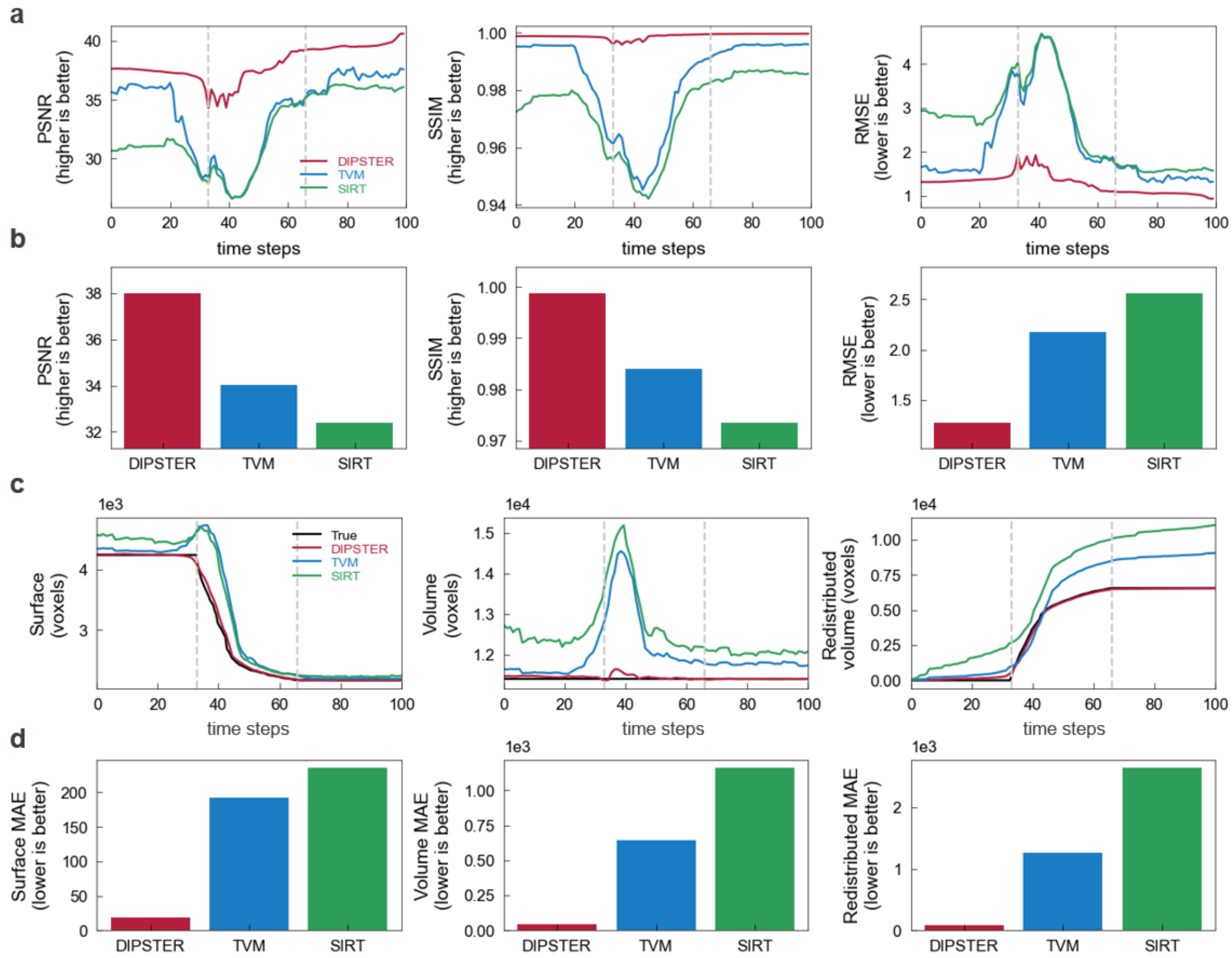

**Supplementary Fig. 6. Complete metrics on the reconstructions of the synthetic nanostar without corruptions.** a) Time-dependent and b) time-averaged quality metrics from DIP-STER (blue), and moving-window TVM (orange) and SIRT (green) reconstructions. c) Time-dependent and d) time-averaged physical metrics from the ground truth simulations (black), DIP-STER (blue), and moving-window TVM (orange) and SIRT (green)

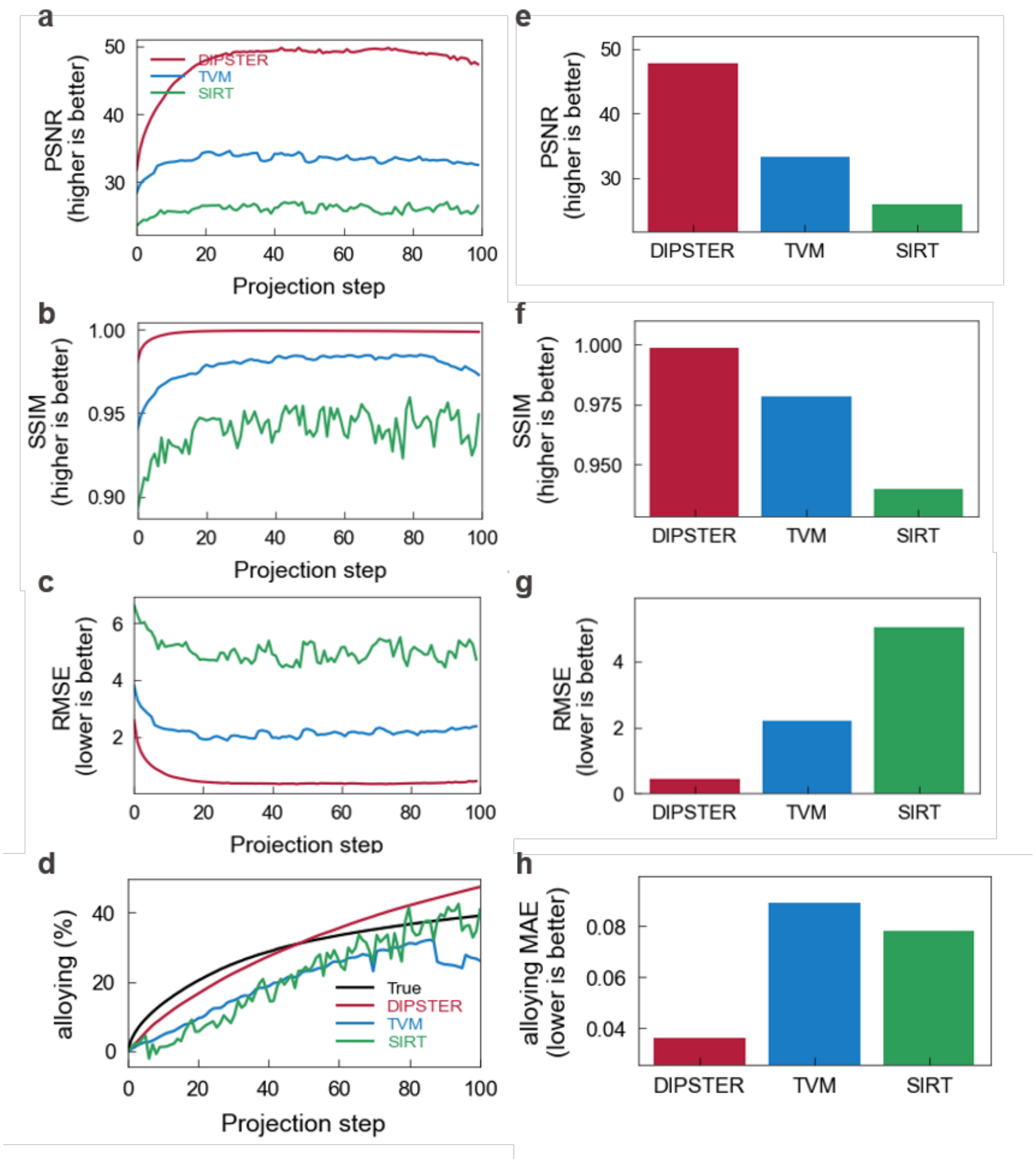

**Supplementary Fig. 7. Complete metrics on the reconstructions of the Au@Ag nanorod without corruptions.** a) Time-dependent and b) time-averaged quality metrics from DIP-STER (blue), moving-window TVM (orange) and SIRT (green) reconstructions. c) Time-dependent and d) time-averaged physical metrics from the ground truth simulations (black), DIP-STER (blue), and moving-window TVM (orange) and SIRT (green)

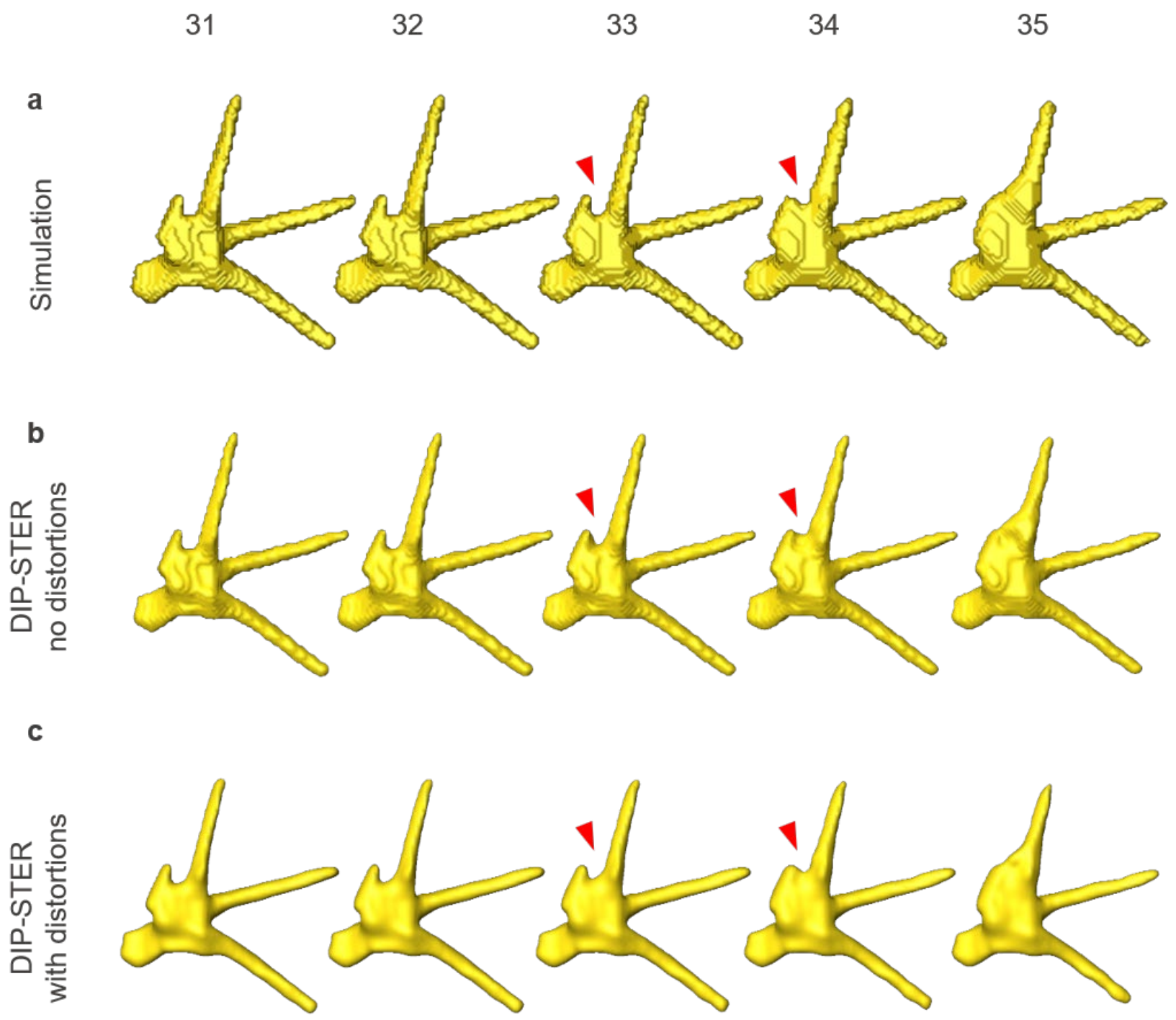

**Supplementary Fig. 8: Localized and abrupt dynamics in simulations and DIP-STER reconstructions**. a) Simulation of melting dynamics with degradation onset at frame 34. b) DIP-STER reconstructions from pristine and c) corrupted tiltseries. Arrows are guides to the eye showing the initial redistribution effects. Both DIP-STER series accurately retrieve the onset at the correct frame.

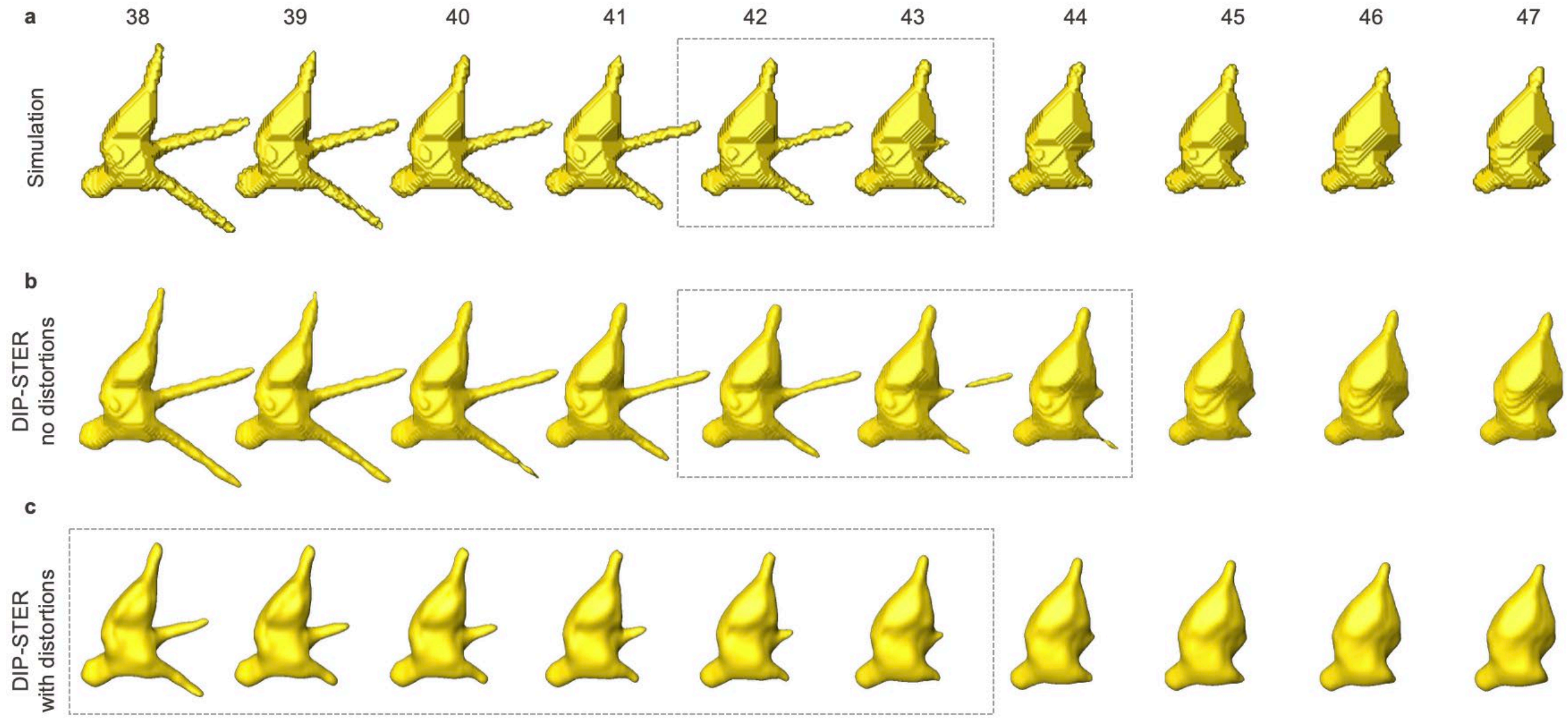

**Supplementary Fig. 9: Localized and abrupt dynamics in simulations and DIP-STER reconstructions**. a) Simulation of melting dynamics with occasional branch loss between frames. b) DIP-STER reconstructions from pristine and c) corrupted tiltseries. The dashed boxes indicate the frames through which the branch loss event spreads.

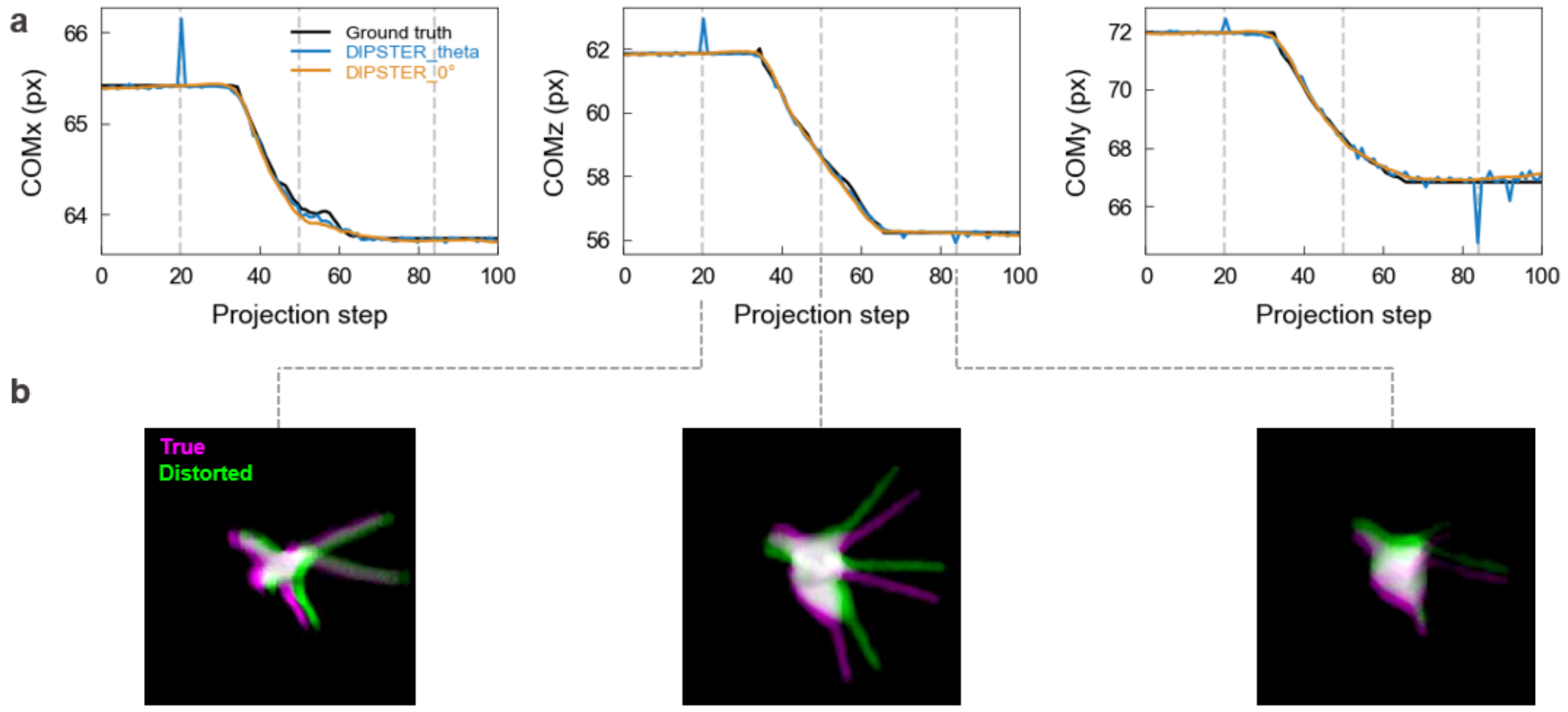

**Supplementary Fig. 10. Effect of outlier projections on DIP-STER.** a) Center of mass (COM) of ground truth volumes (black), DIP-STER volumes from the ($y_k$, $t_i$, $\theta_i$) manifold coordinates (blue) and DIP-STER volumes from the ($y_k$, $t_i$, $\theta = 0°$) manifold coordinates (orange). b) Translation and rotation outlier introduced at the dashed times.

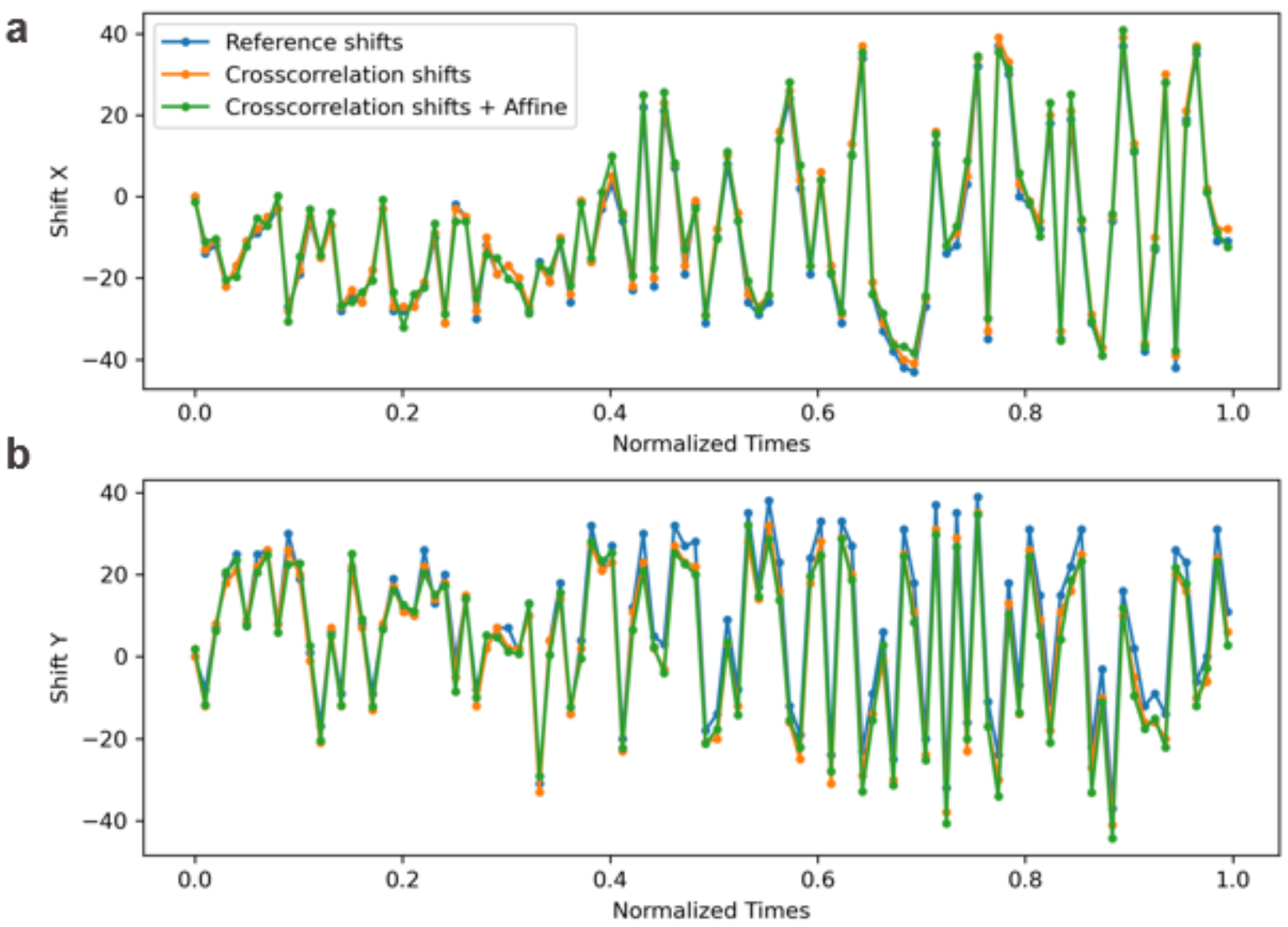

**Supplementary Fig. 11. Shifts retrieved by cross-correlation.** a) x- and b) y-shifts retrieved by cross-correlation on corrupted tiltseries with random translation. Including a model for affine transformation during DIP-STER training did not significantly improved the results.

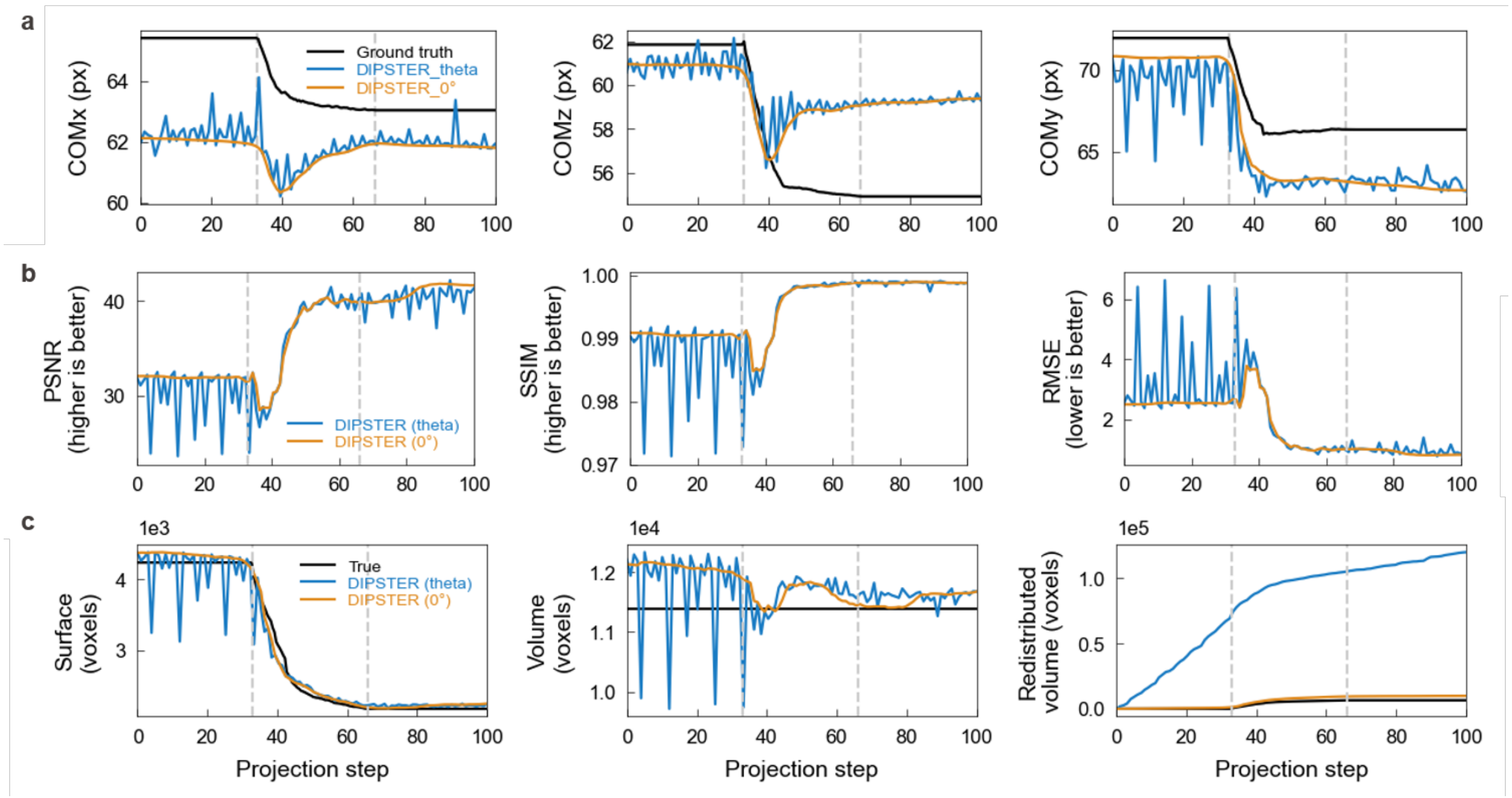

**Supplementary Fig. 12. Effect of residual distortions on DIP-STER** estimated from the synthetic nanostar series with experimental corruption. a) Position of the COM of ground truth volumes (black), DIP-STER volumes from the ($y_k$, $t_i$, $\theta_i$) manifold coordinates (blue) and DIP-STER volumes from the ($y_k$, $t_i$, $\theta = 0°$) manifold coordinates (orange). The DIP-STER reconstructions used tiltseries simulated with experimentally based corruptions, including Poisson and scan noise, tilt-dependent background intensities and random translations. They were processed with the same pipeline as the experimental nanostar before training. b) Quality metrics and c) physical metrics on the same volume-time series.

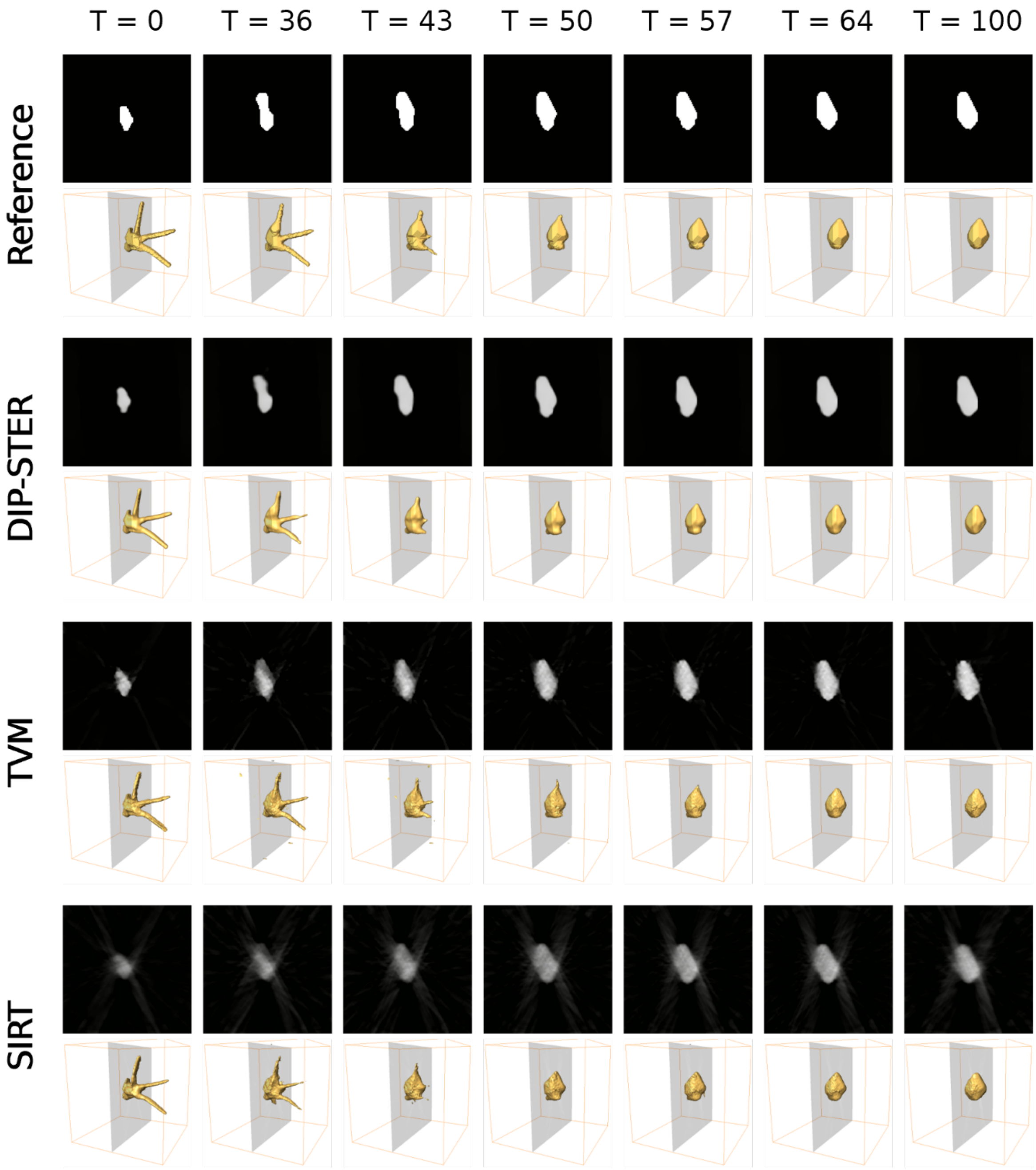

**Supplementary Fig. 13. Comparison of the reconstruction of a distorted simulated nanostar** with redistribution of the volume while heating with DIP-STER, TVM and SIRT. Here again DIP-STER reconstruction are the closest to the ground truth volume. SIRT and TVM have a rougher volume surface and light artefacts in the orthoslices.

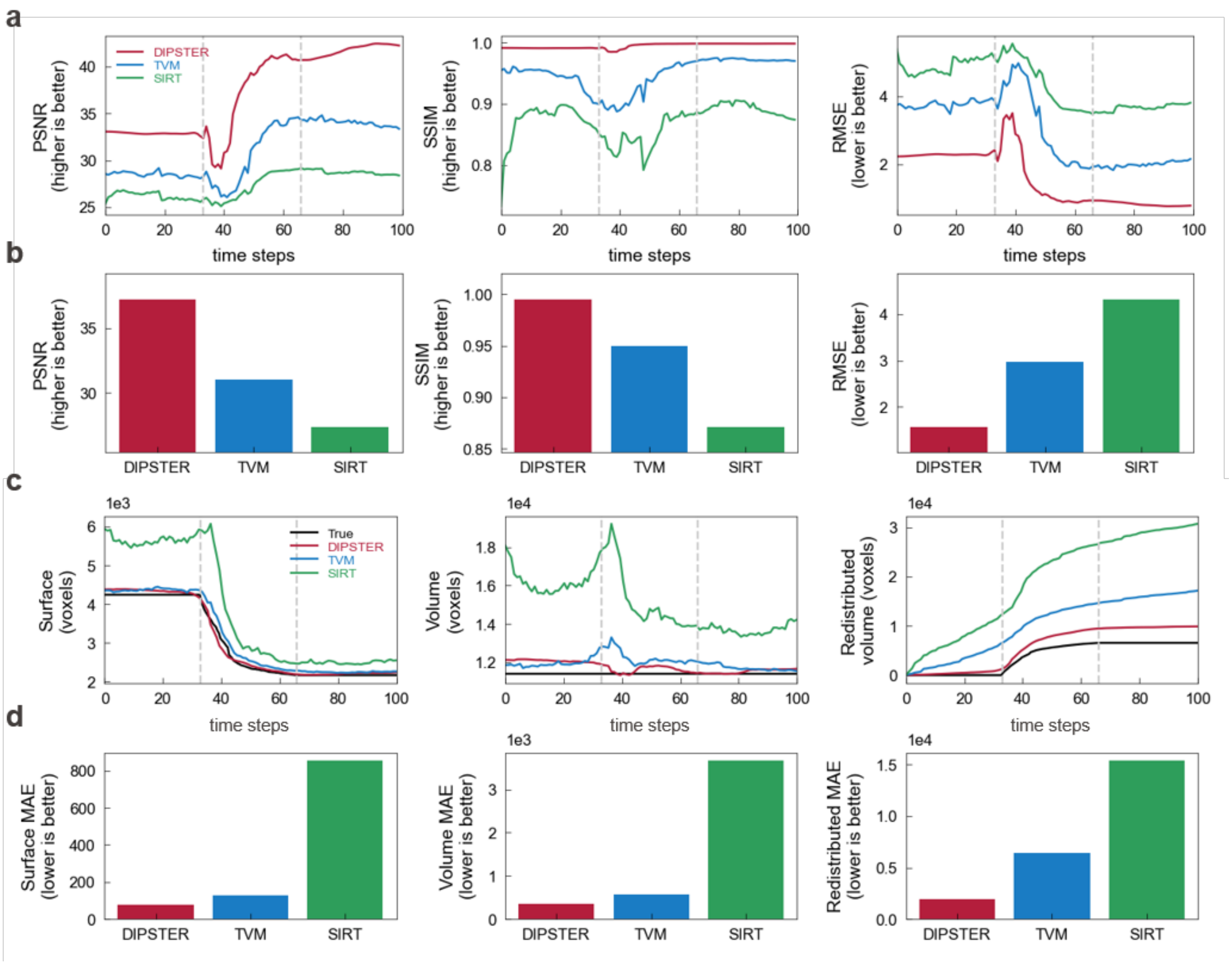

**Supplementary Fig. 14. Additional metrics on the reconstructions of the synthetic nanostar with corruptions.** a) Time-dependent and b) time-averaged quality metrics from DIPSTER (blue), and moving-window TVM (orange) and SIRT (green) reconstructions. c) Time-dependent and d) time-averaged physical metrics from the ground truth simulations (black), DIPSTER (blue), and moving-window TVM (orange) and SIRT (green)

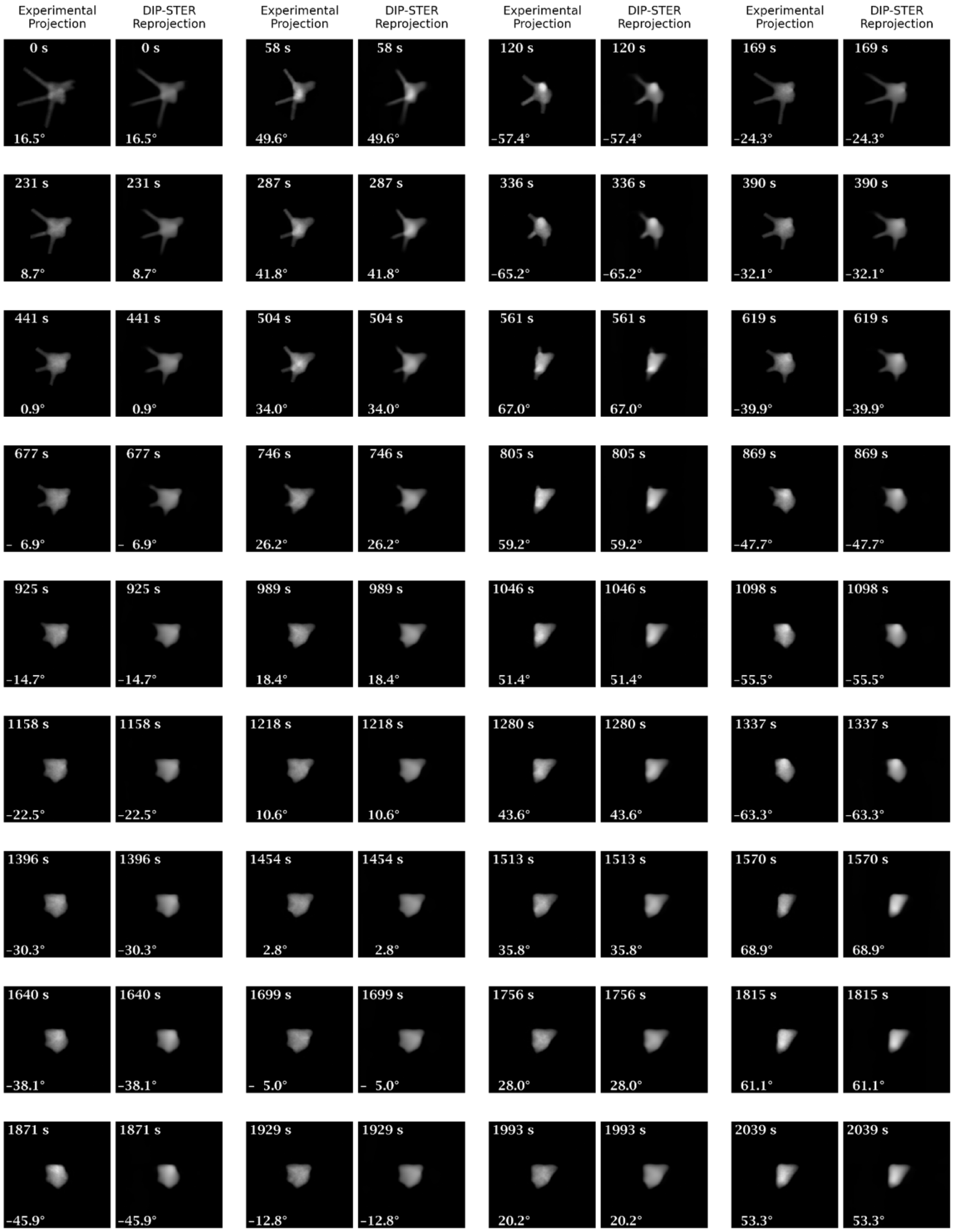

**Supplementary Fig 15. Comparison of the experimental tilt series of a gold nanostar** (left columns) with its forward projected DIP-STER reconstructions (right columns). Only half of the frames of the tilt series are displayed. There is a good agreement in the reprojection with DIP-STER, even for the fine branches.

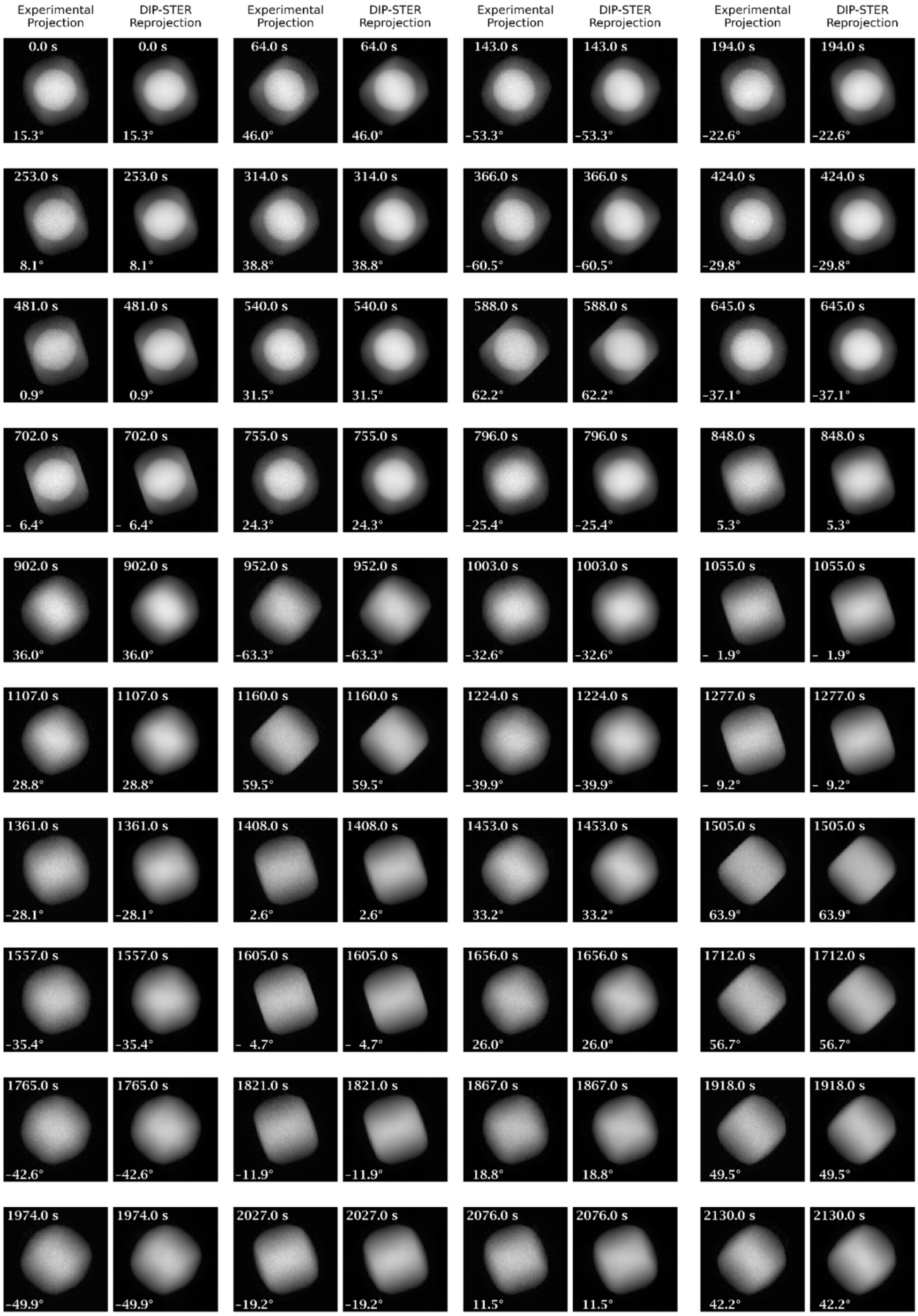

**Supplementary Fig 16. Comparison of the experimental tilt series of an Au@Ag nanocube** (left columns) with its forward projected DIP-STER reconstruction (right columns). Only half of the frames of the tilt series are displayed. There is a good agreement in the reprojection with DIP-STER, with noticeable denoising.

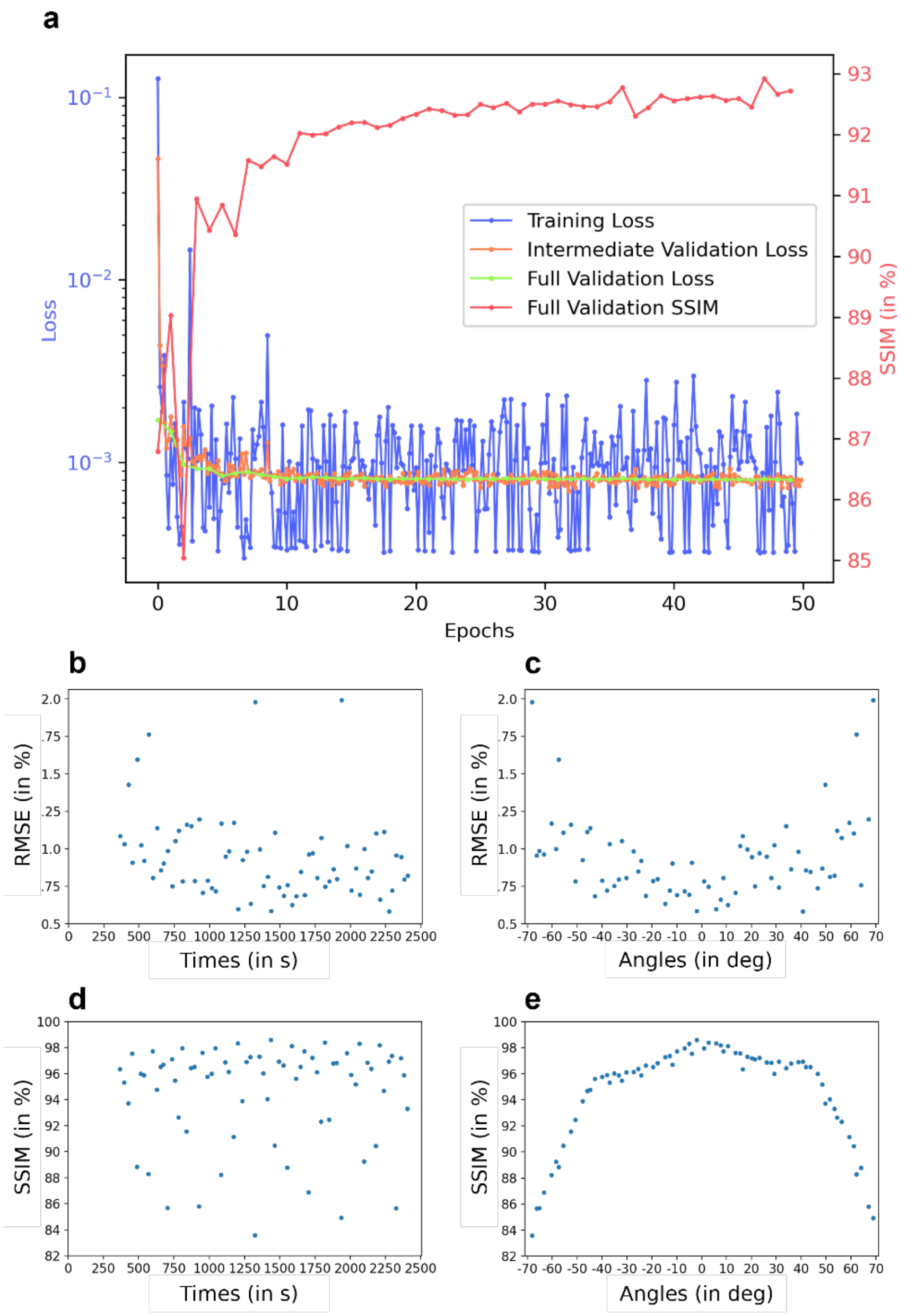

**Supplementary Fig 17. Training evaluation for the gold nanostar**. a) Training curve for the gold nanostar. In blue is represented the evolution of the loss (Geman-McClure) as a function of the number of steps and in red the evolution of the SSIM (our evaluation metric) for the same steps. The loss shows strong fluctuations between different steps, likely due to the presence of noise, as the less noise-sensitive SSIM appear more stable. Two validation strategies are implemented; full validation measurements at the end of every epoch with the entire validation set, and intermediate measurements on a subset of the validation split monitored 4-8 times per epoch. b), c) RMSE and d) ,e) SSIM measurements after training, as function of b), d) the time and c), e) the angle of projection acquisition. The metrics were calculated between experimental projections and DIP-STER reprojections. The metrics remain uncorrelated to the time of acquisition, but the SSIM shows a noticeable decrease at the high tilt angle. This effect could be related to an imperfect correction of the detector shadowing despite the intensity normalization steps described in the main **Methods**.

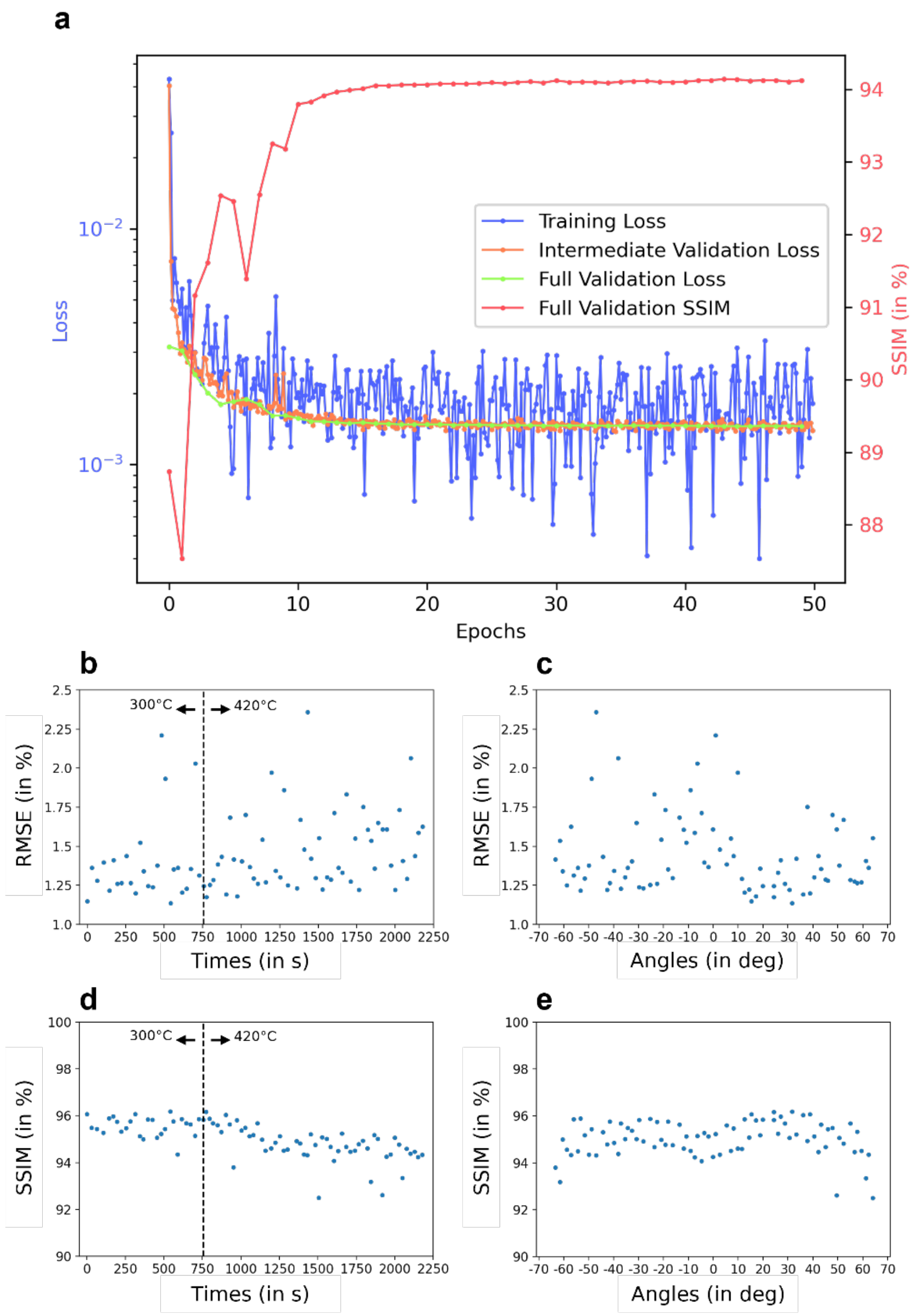

**Supplementary Fig 18. Training evaluation for the Au@Ag nanocube**. (a) Training and validation curves for the Au@Ag nanocube. In blue is represented the evolution of the loss as a function of the number of steps and in red the evolution of the SSIM for the same steps. Compared to the nanostar we stopped earlier as the loss is more stable and denoising is improved. b), c) RMSE and d), e) SSIM measurements after training, as function of b), d) the time and c), e) the angle of projection acquisition. The metrics were calculated between experimental projections and DIP-STER reprojections. Similar to the nanostar, the metrics remain uncorrelated to the time of acquisition, but the SSIM shows a less noticeable decrease at high tilt angles.

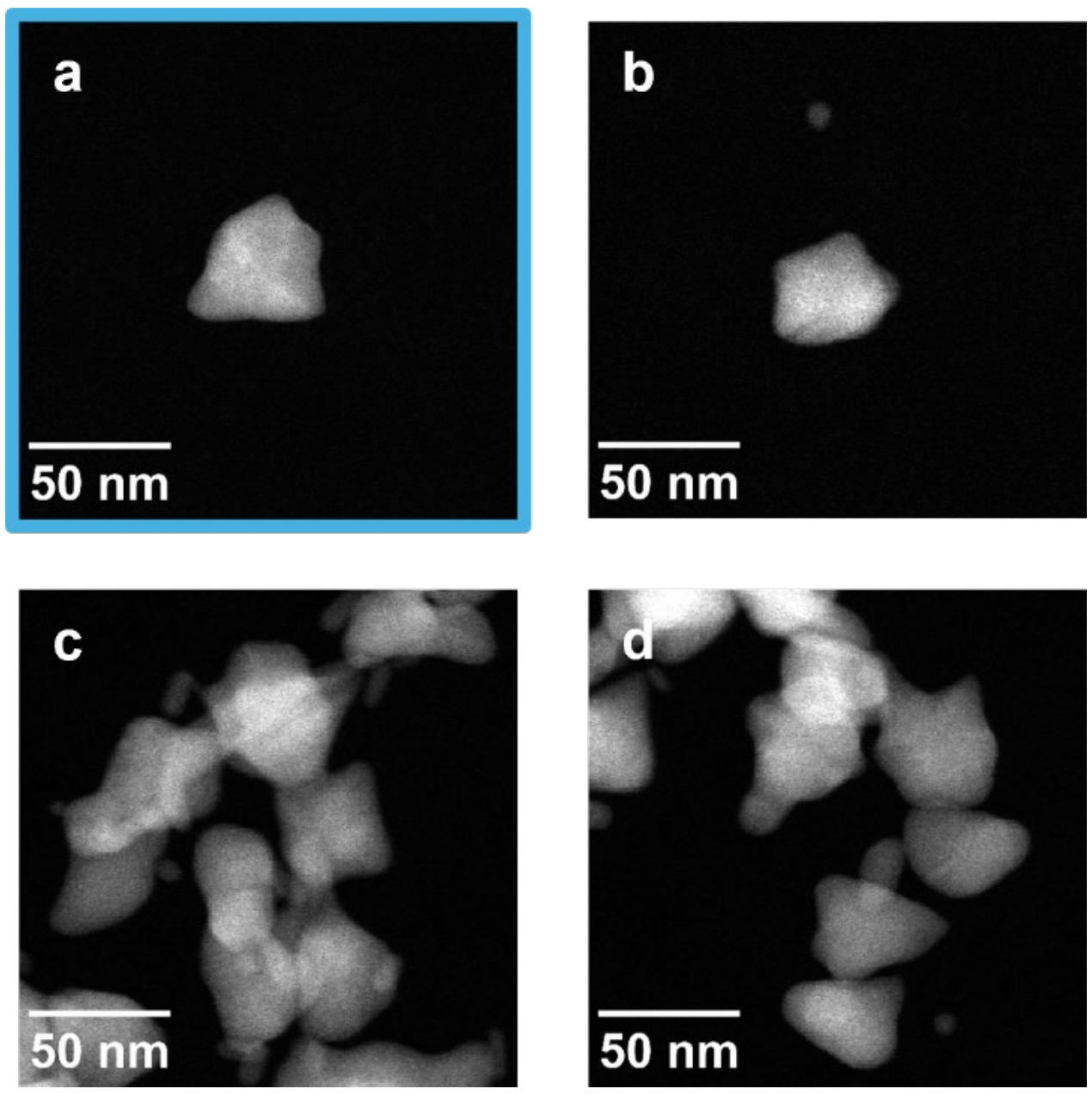

**Supplementary Fig 19. Comparison of heated Au nanostars outside of the field of view**. a) Final state of the experimental nanostar tracked for DIP-STER with a constant illumination during the tilt series acquisition (blue frame). b) Isolated nanostar. c-d) Assembly of nanostars. The chip was heated at 220 °C for a total of 35 minutes. Regardless of their isolated or aggregated state, all particles degraded similarly to the one exposed to the e-beam. This demonstrates that, unlike previous reports,[18,25–27] the dynamics were not modified by the e-beam likely due to a lower accumulated electron dose (**Supplementary Note 3**).

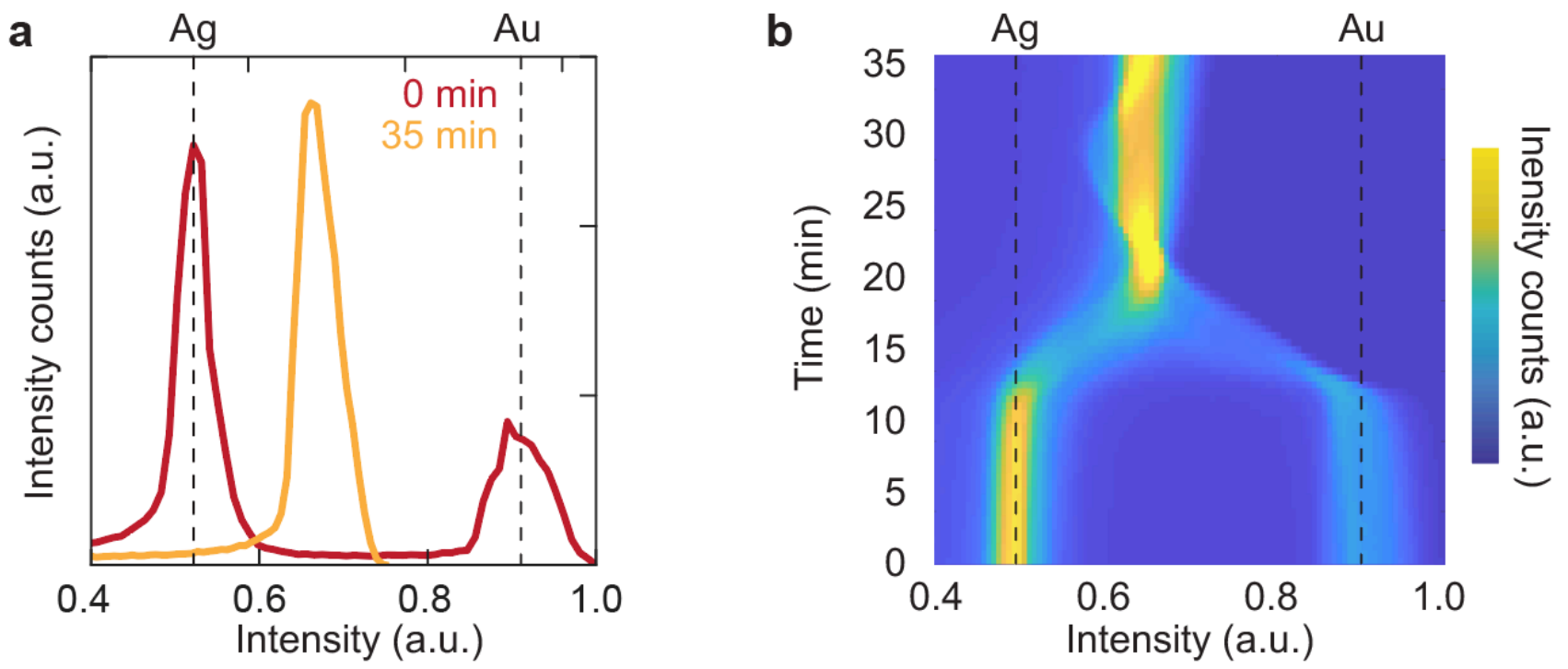

**Supplementary Fig 20. Intensity evolution in the Au@Ag nanocube reconstructed with DIP-STER**. a) histogram of the 3D reconstructions at the start and end of the volume-time series and b) 2D histogram of the volume time-series. The clear peaks at the start are associated with the core-shell Au@Ag structure, and gradually blend as diffusion and alloying progress. This reduces the variance in the histogram, which was used to quantify the alloying progress as described in the main **Methods**.

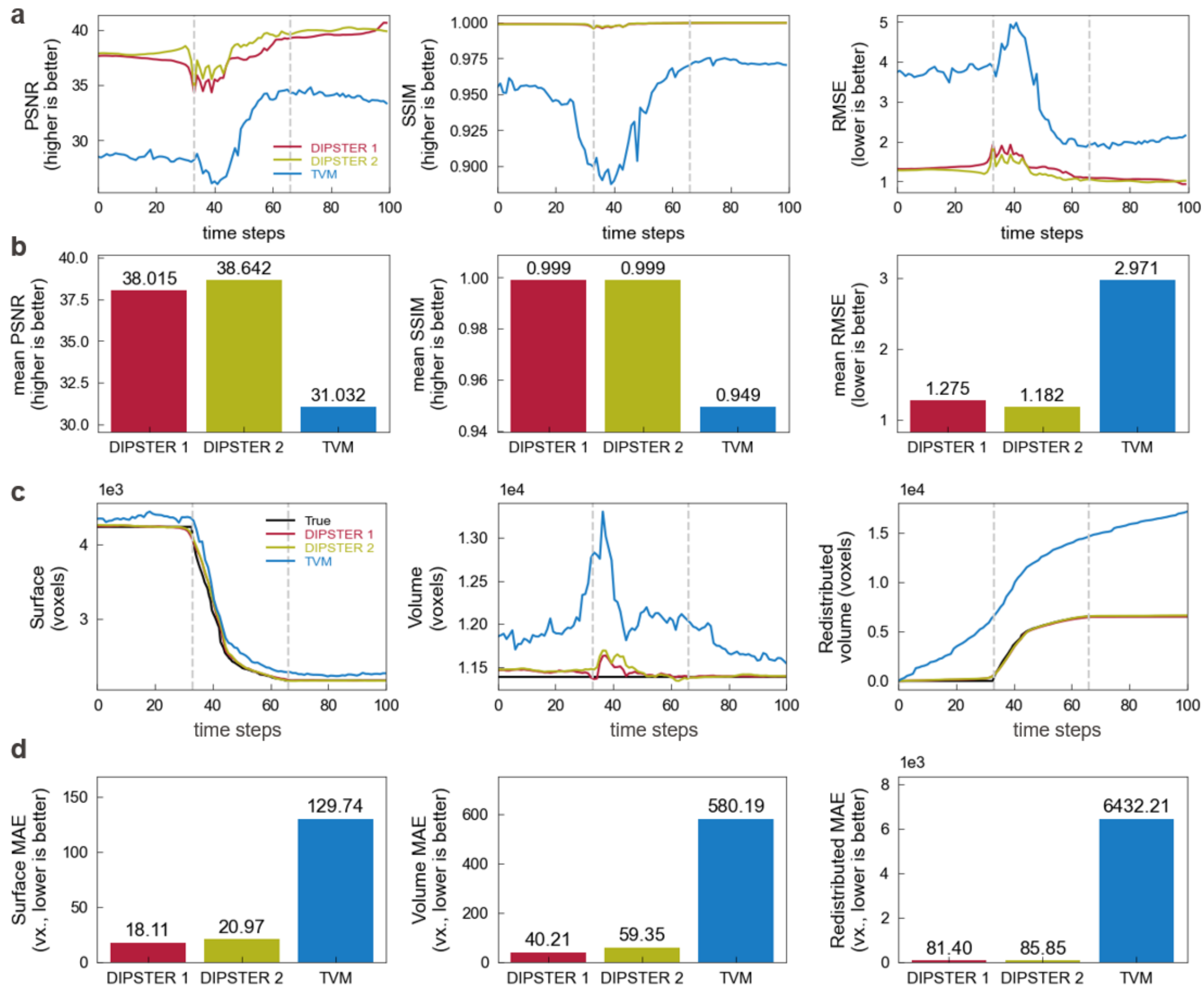

**Supplementary Fig 21. DIP-STER reproducibility**. a) Comparison the time-dependant and b) time average quality metrics of two DIP-STER network trained on an un-degraded tilt series of the melting nanostar with identical hyperparameters. c) Comparison of the time-dependant and d) time average physical metrics. TVM results are reproduced from **Supplementary Fig. 6** for reference. The run-on-run difference appears small across all metrics, despite the stochastic nature of network initialization.

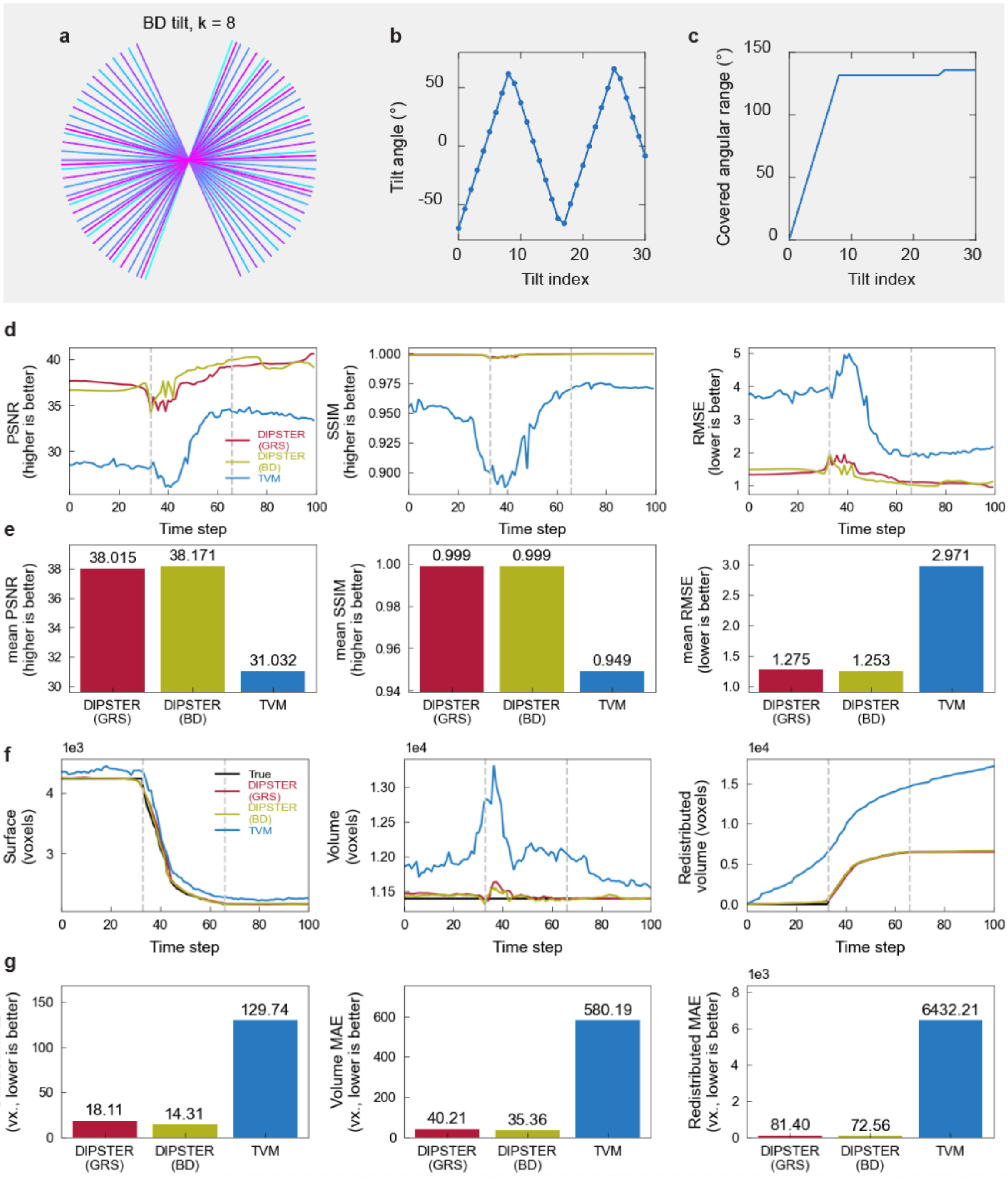

**Supplementary Fig. 22. Comparison of continuous acquisition schemes on the simulated nanostar**. a) Visualization of the binary decomposition (BD) tilting scheme as described in ref.[15], b) tilt angles and c) covered angular range for a tilt series of 30 images. d) Comparison the time-dependant and e) time average quality metrics of DIP-STER trained on (un-degraded) BD or GRS tilt series of the melting nanostar against the ground truth simulations. f) Comparison of  the time-dependant and e) time average physical metrics. TVM results are reproduced from **Supplementary Fig. 6** for reference. It was experimentally evaluated that the smaller angular steps with BD afforded ~2x more frames compared to GRS; about 10-15 s/frame for BD, 20-30 s/frame for GRS. Training was therefore performed with 100 GRS frames and 200 BD frames for the same overall time. Both schemes appear to perform very similarly with results within DIP-STER's variability.

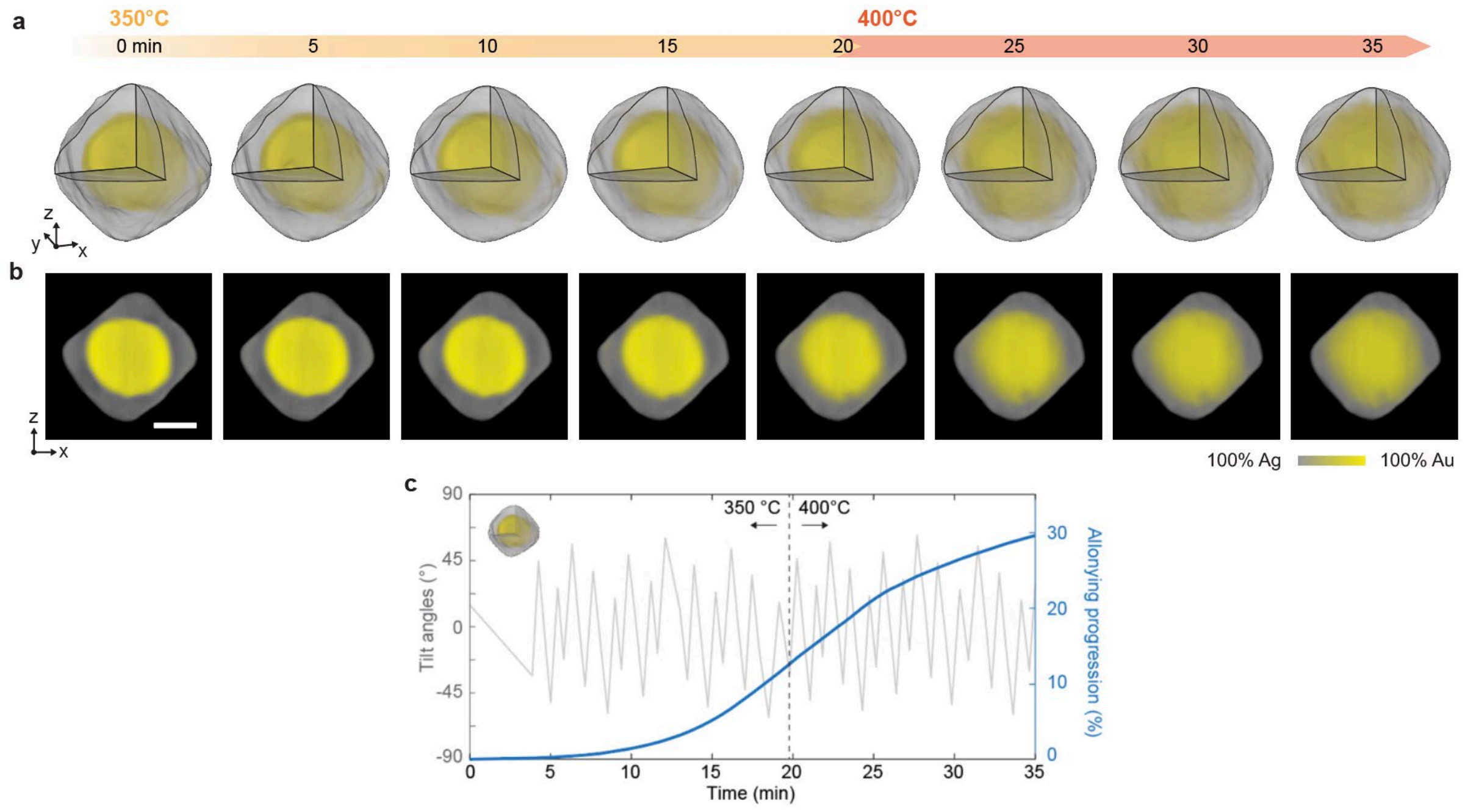

**Supplementary Fig. 23. Experimental results on an alloying Au@Ag nanocube with strong distortions**. a) Volume rendering and b) orthoslice of DIP-STER volumes of a silica-coated Au@Ag core-shell nanocube heated to 350°C then 400°C. Scale bar is 20 nm. c) Experimental projection angles in the Au@Ag tilt series as a function of time, and alloying progress over time, as estimated from the spread of intensities in the reconstructions. This tilt series contained stronger scan distortions than the one shown in **Fig. 5** of the main text, and was recorded in the ±70° range with noticeable drops in intensity above 60° (**Supplementary Fig. 24**). Although the temperature step was smaller than that of **Fig. 5**, with some alloying already expected at 350°C, a sharper step in the alloying progression would be expected. This suggests again that a lower quality tilt series reduces the temporal resolution of DIP-STER series.

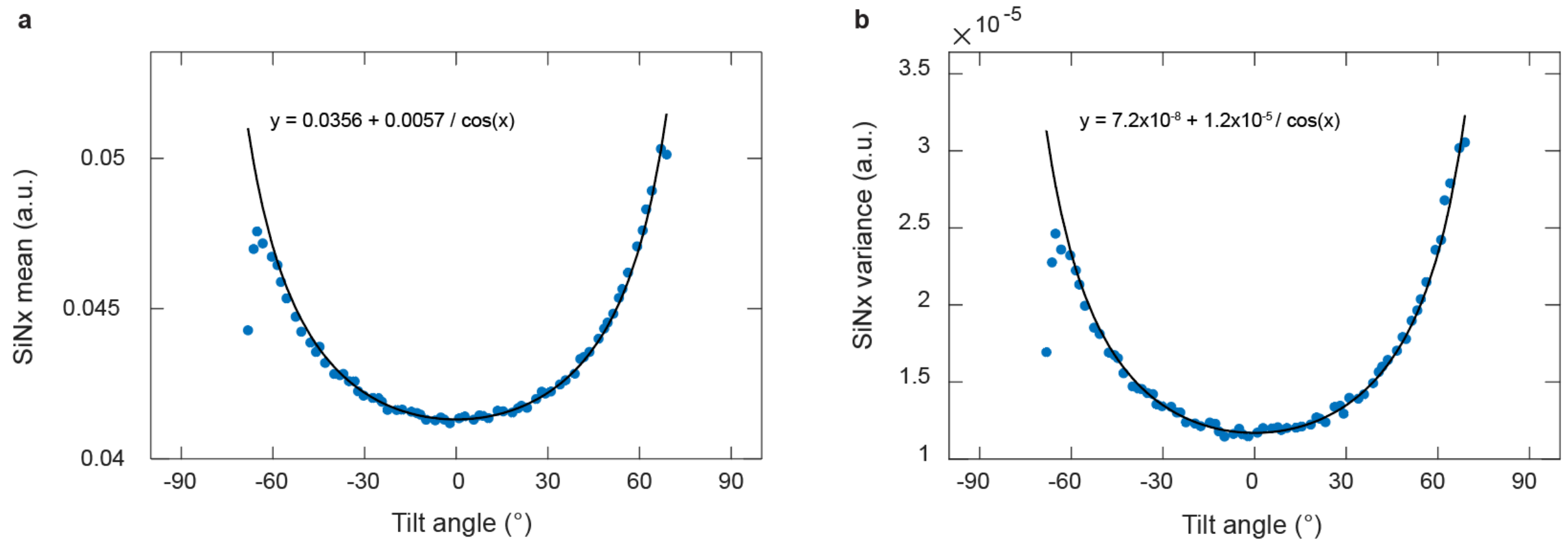

**Supplementary Fig. 24. Detector shadowing at high tilt angles**. a) Change in the mean and b) variance of the intensity distribution in an area containing only the SiNx membrane as a function of the tilt angle for the tilt series of the Au@Ag nanocube in **Supplementary Fig. 23**. The mean and the variance follow the same trend, consistently with the STEM signal being a Poisson process. At most angle, the trend is accurately modelled by a function proportional to 1/cos(θ), with θ the tilt angle, corresponding to an increase in membrane thickness in the e-beam path. However, high angles (typically > 60°) deviate from the model, likely due to partial shadowing of the HAADF detector by the holder. The exact angle at which shadowing will start occurring depends on the eucentric height and sample position on the grid, and therefore of a specific holder-microscope configuration. For the data in **Fig. 5** of the main text, we confirmed that the 1/cos(θ) trend was accurately followed up to ±65°.

## References

1. Jenkinson, K., Liz-Marzán, L. M. & Bals, S. Multimode Electron Tomography Sheds Light on Synthesis, Structure, and Properties of Complex Metal-Based Nanoparticles. *Adv. Mater.* **34**, 2110394 (2022).
2. Weyland, M. Electron Tomography of Catalysts. *Top. Catal.* **21**, 175–183 (2002).
3. Midgley, P. A. & Dunin-Borkowski, R. E. Electron tomography and holography in materials science. *Nat. Mater.* **8**, 271–280 (2009).
4. Goris, B. *et al.* Atomic-scale determination of surface facets in gold nanorods. *Nat. Mater.* **11**, 930–935 (2012).
5. Goris, B., Roelandts, T., Batenburg, K. J., Mezerji, H. H. & Bals, S. Advanced reconstruction algorithms for electron tomography: from comparison to combination. *Ultramicroscopy* **127**, 40–47 (2013).
6. Chen, D. *et al.* The properties of SIRT, TVM, and DART for 3D imaging of tubular domains in nanocomposite thin-films and sections. *Ultramicroscopy* **147**, 137–148 (2014).
7. Egerton, R. F. Radiation damage to organic and inorganic specimens in the TEM. *Micron* **119**, 72–87 (2019).
8. Egerton, R. F. Mechanisms of radiation damage in beam-sensitive specimens, for TEM accelerating voltages between 10 and 300 kV. *Microsc. Res. Tech.* **75**, 1550–1556 (2012).
9. Chee, S. W., Lunkenbein, T., Schlögl, R. & Cuenya, B. R. In situ and operando electron microscopy in heterogeneous catalysis—insights into multi-scale chemical dynamics. *J. Phys. Condens. Matter* **33**, 153001 (2021).
10. Vanrompay, H. *et al.* 3D characterization of heat-induced morphological changes of Au nanostars by fast: In situ electron tomography. *Nanoscale* **10**, 22792–22801 (2018).
11. Ko, K. *et al.* Operando electron microscopy investigation of polar domain dynamics in twisted van der Waals homobilayers. *Nat. Mater.* **22**, 992–998 (2023).
12. Chee, S. W., Lunkenbein, T., Schlögl, R. & Roldán Cuenya, B. Operando Electron Microscopy of Catalysts: The Missing Cornerstone in Heterogeneous Catalysis Research? *Chem. Rev.* **123**, 13374–13418 (2023).
13. Yang, R. *et al.* Fabrication of liquid cell for in situ transmission electron microscopy of electrochemical processes. *Nat. Protoc.* **18**, 555–578 (2023).
14. Lee, S., Schneider, N. M., Tan, S. F. & Ross, F. M. Temperature Dependent Nanochemistry and Growth Kinetics Using Liquid Cell Transmission Electron Microscopy. *ACS Nano* **17**, 5609–5619 (2023).
15. Craig, T. M., Girod, R., Vinnacombe-Willson, G., Liz-Marzán, L. M. & Bals, S. Towards continuous time-dependent tomography: implementation and evaluation of continuous acquisition schemes in electron tomography. *Ultramicroscopy* **277**, 114207 (2025).
16. Albrecht, W. & Bals, S. Fast Electron Tomography for Nanomaterials. *J. Phys. Chem. C* **124**, 27276–27286 (2020).
17. Vanrompay, H. *et al.* Fast versus conventional HAADF-STEM tomography of nanoparticles: advantages and challenges. *Ultramicroscopy* **221**, (2021).
18. Vanrompay, H. *et al.* 3D characterization of heat-induced morphological changes of Au nanostars by fast in situ electron tomography. *Nanoscale* **10**, 22792–22801 (2018).
19. Eisenstein, F., Danev, R. & Pilhofer, M. Improved applicability and robustness of fast cryo-electron tomography data acquisition. *J. Struct. Biol.* **208**, 107–114 (2019).
20. Chreifi, G., Chen, S., Metskas, L. A., Kaplan, M. & Jensen, G. J. Rapid tilt-series acquisition for electron cryotomography. *J. Struct. Biol.* **205**, 163–169 (2019).
21. Albrecht, W., Van Aert, S. & Bals, S. Three-Dimensional Nanoparticle Transformations Captured by an Electron Microscope. *Acc. Chem. Res.* **54**, 1189–1199 (2021).
22. Mychinko, M. *et al.* The Influence of Size, Shape, and Twin Boundaries on Heat-Induced Alloying in Individual Au@Ag Core–Shell Nanoparticles. *Small* **17**, 2102348 (2021).

23. Skorikov, A. *et al.* Quantitative 3D Characterization of Elemental Diffusion Dynamics in Individual Ag@Au Nanoparticles with Different Shapes. *ACS Nano* **13**, 13421–13429 (2019).
24. Albrecht, W. *et al.* Thermal stability of gold/palladium octopods studied in situ in 3D: Understanding design rules for thermally stable metal nanoparticles. *ACS Nano* **13**, 6522–6530 (2019).
25. De Meyer, R., Albrecht, W. & Bals, S. Effectiveness of reducing the influence of CTAB at the surface of metal nanoparticles during in situ heating studies by TEM. *Micron* **144**, 103036 (2021).
26. Albrecht, W. *et al.* Thermal Stability of Gold/Palladium Octopods Studied *in Situ* in 3D: Understanding Design Rules for Thermally Stable Metal Nanoparticles. *ACS Nano* **13**, 6522–6530 (2019).
27. Albrecht, W. *et al.* Impact of the electron beam on the thermal stability of gold nanorods studied by environmental transmission electron microscopy. *Ultramicroscopy* **193**, 97–103 (2018).
28. Münch, B. Spatiotemporal computed tomography of dynamic processes. *Opt. Eng.* **50**, 123201 (2011).
29. Craig, T. M., Kadu, A. A., Batenburg, K. J. & Bals, S. Real-time tilt undersampling optimization during electron tomography of beam sensitive samples using golden ratio scanning and RECAST3D. *Nanoscale* **15**, 5391–5402 (2023).
30. Batenburg, K. J. & Sijbers, J. DART: A Practical Reconstruction Algorithm for Discrete Tomography. *IEEE Trans. Image Process.* **20**, 2542–2553 (2011).
31. Leary, R., Saghi, Z., Midgley, P. A. & Holland, D. J. Compressed sensing electron tomography. *Ultramicroscopy* **131**, 70–91 (2013).
32. Arenas Esteban, D. *et al.* Quantitative 3D structural analysis of small colloidal assemblies under native conditions by liquid-cell fast electron tomography. *Nat. Commun.* **15**, 6399 (2024).
33. Lobato, I., Friedrich, T. & Van Aert, S. Deep convolutional neural networks to restore single-shot electron microscopy images. *Npj Comput. Mater.* **10**, 10 (2024).
34. Girod, R., Lazaridis, T., Gasteiger, H. A. & Tileli, V. Three-dimensional nanoimaging of fuel cell catalyst layers. *Nat. Catal.* **6**, 383–391 (2023).
35. Buchholz, T.-O., Jordan, M., Pigino, G. & Jug, F. Cryo-CARE: Content-Aware Image Restoration for Cryo-Transmission Electron Microscopy Data. in *2019 IEEE 16th International Symposium on Biomedical Imaging (ISBI 2019)* 502–506 (2019). doi:10.1109/ISBI.2019.8759519.
36. Yao, L., Lyu, Z., Li, J. & Chen, Q. No ground truth needed: unsupervised sinogram inpainting for nanoparticle electron tomography (UsiNet) to correct missing wedges. *Npj Comput. Mater.* **10**, 28 (2024).
37. Liu, Y.-T. *et al.* Isotropic reconstruction for electron tomography with deep learning. *Nat. Commun.* **13**, 6482 (2022).
38. Zhang, Y., Cao, L., Sun, Z. & Zhou, J. Physics-aware neural networks enable robust and full atomic structure determination via low-dose atomic electron tomography. Preprint at https://arxiv.org/abs/2603.19942v1 (2026).
39. Lee, J. & Yang, Y. Advancing atomic electron tomography with neural networks. *Appl. Microsc.* **55**, 7 (2025).
40. Moncomble, A. *et al.* aquaDenoising: AI-enhancement of in situ liquid phase STEM video for automated quantification of nanoparticles growth. *Ultramicroscopy* **271**, 114121 (2025).
41. Kim, J. *et al.* Self-supervised machine learning framework for high-throughput electron microscopy. *Sci. Adv.* **11**, eads5552 (2025).
42. Crozier, P. A. *et al.* Visualizing nanoparticle surface dynamics and instabilities enabled by deep denoising. *Science* **387**, 949–954 (2025).

43. Pelaez-Fernandez, M. *et al.* Low-Dose 3D Bonding Mapping Through 'Soft' Core-Loss EELS Tomography and Unsupervised Deep Learning. Preprint at https://doi.org/10.48550/arXiv.2606.10893 (2026).
44. Brosset, S. *et al.* Unsupervised deep image prior for sparse-view and limited-angle electron tomography. *Ultramicroscopy* **285**, 114414 (2026).
45. Lim, C. *et al.* Missing Wedge Inpainting and Joint Alignment in Electron Tomography through Implicit Neural Representations. Preprint at https://doi.org/10.48550/arXiv.2512.08113 (2025).
46. Yoo, J. *et al.* Time-Dependent Deep Image Prior for Dynamic MRI. *IEEE Trans. Med. Imaging* **40**, 3337–3348 (2021).
47. Reed, A. W. *et al.* Dynamic CT Reconstruction from Limited Views with Implicit Neural Representations and Parametric Motion Fields. in *2021 IEEE/CVF International Conference on Computer Vision (ICCV)* 2238–2248 (2021). doi:10.1109/ICCV48922.2021.00226.
48. Zhang, Y., Shao, H.-C., Pan, T. & Mengke, T. Dynamic cone-beam CT reconstruction using spatial and temporal implicit neural representation learning (STINR). *Phys. Med. Biol.* **68**, 045005 (2023).
49. Zhang, Y., Yao, Z., Klöfkorn, R., Ritschel, T. & Villanueva-Perez, P. 4D-ONIX for reconstructing 3D movies from sparse X-ray projections via deep learning. *Commun. Eng.* **4**, 54 (2025).
50. Boulanger, M. & Jossou, E. Interlaced dynamic XCT reconstruction with spatio-temporal implicit neural representations. Preprint at https://doi.org/10.48550/arXiv.2510.08641 (2025).
51. Chien, T., Ophus, C. & Waller, L. Space-Time Implicit Neural Representations for Atomic Electron Tomography on Dynamic Samples. in (2023).
52. Ye, E. *et al.* Recovering a Molecule's 3D Dynamics from Liquid-phase Electron Microscopy Movies. Preprint at https://doi.org/10.48550/ARXIV.2308.11927 (2023).
53. Ulyanov, D., Vedaldi, A. & Lempitsky, V. Deep Image Prior. *Int. J. Comput. Vis.* **128**, 1867–1888 (2020).
54. Belkin, M., Niyogi, P. & Sindhwani, V. Manifold Regularization: A Geometric Framework for Learning from Labeled and Unlabeled Examples. *J. Mach. Learn. Res.* **7**, 2399–2434 (2006).
55. Nakarmi, U., Slavakis, K., Lyu, J. & Ying, L. M-MRI: A manifold-based framework to highly accelerated dynamic magnetic resonance imaging. in *Proc. IEEE 14th Int. Symp. Biomed. Imag.* 19–22 (2017).
56. Barutcu, S., Aslan, S., Katsaggelos, A. K. & Gürsoy, D. Limited-angle computed tomography with deep image and physics priors. *Sci. Rep.* **11**, 17740 (2021).
57. Baguer, D. O., Leuschner, J. & Schmidt, M. Computed tomography reconstruction using deep image prior and learned reconstruction methods. *Inverse Probl.* **36**, 094004 (2020).
58. Çiçek, Ö., Abdulkadir, A., Lienkamp, S. S., Brox, T. & Ronneberger, O. 3D U-Net: Learning Dense Volumetric Segmentation from Sparse Annotation. in *Medical Image Computing and Computer-Assisted Intervention – MICCAI 2016* 424–432 (Cham, 2016).
59. Barron, J. T. A general and adaptive robust loss function. in *Proc. IEEE/CVF Conf. Comput. Vis. Pattern Recognit.* 4326–4334 (2019).
60. Liu, J., Sun, Y., Xu, X. & Kamilov, U. S. Image Restoration Using Total Variation Regularized Deep Image Prior. in *ICASSP 2019 - 2019 IEEE International Conference on Acoustics, Speech and Signal Processing (ICASSP)* 7715–7719 (2019).
61. van der Hoeven, J. E. S. *et al.* In Situ Observation of Atomic Redistribution in Alloying Gold–Silver Nanorods. *ACS Nano* **12**, 8467–8476 (2018).
62. Skorikov, A. *et al.* Quantitative 3D Characterization of Elemental Diffusion Dynamics in Individual Ag@Au Nanoparticles with Different Shapes. *ACS Nano* **13**, 13421–13429 (2019).
63. Mychinko, M. *et al.* The Influence of Size, Shape, and Twin Boundaries on Heat-Induced Alloying in Individual Au@Ag Core–Shell Nanoparticles. *Small* **17**, 2102348 (2021).

64. González-Rubio, G. *et al.* Micelle-directed chiral seeded growth on anisotropic gold nanocrystals. *Science* **368**, 1472–1477 (2020).
65. Ophus, C., Ciston, J. & Nelson, C. T. Correcting nonlinear drift distortion of scanning probe and scanning transmission electron microscopies from image pairs with orthogonal scan directions. *Ultramicroscopy* **162**, 1–9 (2016).
66. Chien, T., Ophus, C. & Waller, L. Space-Time Implicit Neural Representations for Atomic Electron Tomography on Dynamic Samples. (2023).
67. Kavak, S. *et al.* High-resolution electron microscopy imaging of MOFs at optimized electron dose. *J. Mater. Chem. A* **13**, 4281–4291 (2025).
68. Skvortsova, I. *et al.* Arm-Length-Controlled CsPbBr3 Nanocrystals for Tunable Optical and Assembly Behavior. *Adv. Mater.* e19211 (2026).
69. Mataev, G., Milanfar, P. & Elad, M. DeepRED: Deep Image Prior Powered by RED. in *Proc. IEEE/CVF Int. Conf. Comput. Vis.* (2019).
70. Madsen, J. & Susi, T. The abTEM code: transmission electron microscopy from first principles. *Open Res. Eur.* **1**, 24 (2021).
71. Atta, S., Beetz, M. & Fabris, L. Understanding the role of AgNO3 concentration and seed morphology in the achievement of tunable shape control in gold nanostars. *Nanoscale* **11**, 2946–2958 (2019).
72. Li, C. *et al.* A simple method to clean ligand contamination on TEM grids. *Ultramicroscopy* **221**, 113195 (2021).
73. Leary, R. K. & Midgley, P. A. Electron Tomography in Materials Science. in *Springer Handbook of Microscopy* (eds Hawkes, P. W. & Spence, J. C. H.) (Springer International Publishing, Cham, 2019).
74. van Aarle, W. *et al.* The ASTRA Toolbox: A platform for advanced algorithm development in electron tomography. *Ultramicroscopy* **157**, 35–47 (2015).
75. van Aarle, W. *et al.* Fast and flexible X-ray tomography using the ASTRA toolbox. *Opt. Express* **24**, 25129 (2016).
76. Hendriksen, A. A. *et al.* Tomosipo: fast, flexible, and convenient 3D tomography for complex scanning geometries in Python. *Opt. Express* **29**, 40494–40513 (2021).
77. Hansen, P. C. & O'Leary, D. P. The Use of the L-Curve in the Regularization of Discrete Ill-Posed Problems. *SIAM J. Sci. Comput.* **14**, 1487–1503 (1993).
78. Schwartz, J. *et al.* Imaging 3D chemistry at 1 nm resolution with fused multi-modal electron tomography. *Nat. Commun.* **15**, 3555 (2024).
79. Akiba, T., Sano, S., Yanase, T., Ohta, T. & Koyama, M. Optuna: A Next-generation Hyperparameter Optimization Framework. in *Proceedings of the 25th ACM SIGKDD International Conference on Knowledge Discovery & Data Mining* 2623–2631 (Association for Computing Machinery, New York, NY, USA, 2019). doi:10.1145/3292500.3330701.
80. Van der Walt, S. *et al.* scikit-image: image processing in Python. *PeerJ* **2**, e453 (2014).
81. Wang, J. *et al.* Anisotropic Total Generalized Variation Enhanced Deep Image Prior for Image Denoising. *Symmetry* **18**, (2026).
82. Temscript 2.1.0. https://temscript.readthedocs.io/stable/about.html.
83. Kaestner, A. P., Munch, B. & Trtik, P. Spatiotemporal computed tomography of dynamic processes. *Opt. Eng.* **50**, 123201 (2011).